\documentclass[a4paper,notoc,nohyper]{JHEP3}
\usepackage{axodraw,psfig}

\def\Re{{\rm Re}}
\def\Li{\mathop{\rm Li}\nolimits}

\def\YEb#1{
\begin{picture}(80, 60)(0, #1)
\Line(10,20)(20, 30)
\Line(10,40)(20,30)
\Line(60,30)(70,40)
\Line(60,30)(70,20)
\CArc(30,30)(10,180,360)
\CArc(50,30)(10,0,360)
\DashLine(30,50)(30,10){4}
\DashCArc(30,30)(10,0,90){1.5}
\SetWidth{2}
\CArc(30,30)(10,90,180)
\end{picture}
}

\def\YAb#1{
\begin{picture}(80,60)(0,#1)
\Line(10,20)(20, 30)
\Line(10,40)(20,30)
\Line(55,50)(65,50)
\Line(55,10)(65,10)
\Line(20,30)(55,10)
\Oval(55,30)(20,5)(0)
\DashLine(37.5,50)(37.5,10){4}
\DashLine(37.5,40)(55, 50){1.5}
\SetWidth{2}
\Line(20, 30)(37.5,40)
\end{picture}
}

\def\YBb#1{
\begin{picture}(80,60)(0,#1)
\Line(10,20)(20, 30)
\Line(10,40)(20,30)
\Line(55,30)(65,40)
\Line(55,30)(65,20)
\CArc(37.5, 30)(17.5, 180, 450)
\CArc(55,47.5)(17.5,180,270)
\DashLine(25,50)(25,10){4}
\DashCArc(37.5, 30)(17.5, 90, 135){1.5}
\SetWidth{2}
\CArc(37.5, 30)(17.5, 135, 180)
\end{picture}
}

\def\YCb#1{
\begin{picture}(80,60)(0,#1)
\Line(10,20)(20, 30)
\Line(10,40)(20,30)
\Line(55,50)(65,50)
\Line(55,10)(65,10)
\Line(20,30)(55,10)
\Line(45,44.28)(55,50)
\Line(45,44.28)(55,10)
\Line(55,50)(50.5,33.2)  \Line(48.3,27)(45,15.72)
\DashLine(32.5,50)(32.5,10){4}
\DashLine(32.5,37.14)(45, 44.28){1.5}
\SetWidth{2}
\Line(20, 30)(32.5,37.14)
\end{picture}
}

\def\YDb#1{
\begin{picture}(80,60)(0,#1)
\Line(10,20)(20, 30)
\Line(10,40)(20,30)
\Line(55,50)(65,50)
\Line(55,10)(65,10)
\Line(20,30)(55,10)
\Line(45,44.28)(55,50)
\Line(45,44.28)(45,15)
\Line(55,50)(55,10)
\DashLine(32.5,50)(32.5,10){4}
\DashLine(32.5,37.14)(45, 44.28){1.5}
\SetWidth{2}
\Line(20, 30)(32.5,37.14)
\end{picture}
}

\def\Xone#1{
\begin{picture}(80,60)(0,#1)
\Line(10,20)(20, 30)
\Line(10,40)(20,30)
\Line(60,30)(70,40)
\Line(60,30)(70,20)
\CArc(40,30)(20,180,360)
\DashLine(40,55)(40,5){4}
\Line(20, 30)(60, 30)
\DashCArc(40,30)(20,0,90){1.5}
\SetWidth{2}
\CArc(40,30)(20,90,180)
\end{picture}
}

\def\Xfourteen#1{
\begin{picture}(80,60)(0,#1)
\Line(10,20)(20, 30)
\Line(10,40)(20,30)
\Line(55,30)(65,40)
\Line(55,30)(65,20)
\CArc(55,12.5)(17.5,90,180)
\Text(23.3,20)[90]{${\bf \times}$}
\CArc(37.5, 30)(17.5,180,270)
\CArc(37.5, 30)(17.5,270,360)
\DashLine(46,50)(46,10){4}
\DashCArc(37.5, 30)(17.5,0,60){1.5}
\SetWidth{2}
\CArc(37.5, 30)(17.5,60,180)
\end{picture}
}

\title{\boldmath 
High-precision QCD at hadron colliders:\\
electroweak gauge boson rapidity distributions\\
at NNLO
\footnote{Research supported by the US Department of Energy under contract 
DE-AC03-76SF00515.}}
\author{
Charalampos Anastasiou$^{a}$,
Lance Dixon$^a$,
Kirill Melnikov$^{b}$ 
and Frank Petriello$^{a,c}$
\\
$^a$ Stanford Linear Accelerator Center, Stanford University, \\
\,\, Stanford, CA 94309, U.S.A.\\
$^b$ Department of Physics \& Astronomy, University of Hawaii, \\
\,\, Honolulu, HI 96822, U.S.A.\\
$^c$ Department of Physics, Johns Hopkins University, 3400 North Charles St., \\ 
\,\, Baltimore, MD 21218, U.S.A. \\
E-mail: \email{babis@slac.stanford.edu}, \email{lance@slac.stanford.edu}, 
\email{kirill@phys.hawaii.edu}, 
\email{frankjp@pha.jhu.edu} }

\abstract{ We compute the rapidity distributions of $W$ and $Z$ bosons
produced at the Tevatron and the LHC through next-to-next-to leading order
in QCD.  Our results demonstrate remarkable stability with respect to
variations of the factorization and renormalization scales for all values
of rapidity accessible in current and future experiments.  These processes
are therefore ``gold-plated'':  current theoretical knowledge 
yields QCD predictions accurate to better than one percent.
These results strengthen the proposal to use $W$ and $Z$ production 
to determine parton-parton luminosities and constrain
parton distribution functions at the LHC.  For example, LHC
data should easily be able to distinguish the central parton 
distribution fit obtained by MRST from that obtained by Alekhin.}

\preprint{{hep-ph/0312266}\\{SLAC--PUB--10288}\\{UH-511-1042-03}\\{December, 2003}}

\begin{document}

\section{Introduction}
\label{sec:intro}

Drell-Yan production of lepton pairs through electroweak (EW) gauge bosons
at hadron colliders occupies a special place in elementary particle
physics.  Historically, the Drell-Yan mechanism~\cite{drellyan} 
was the first application of parton model ideas beyond deep inelastic 
scattering, and was later the route to discovery of the $W$ and 
$Z$ bosons~\cite{WZdiscovery}.
Currently, it provides precise determinations of several Standard Model
(SM) parameters, and places stringent constraints on many forms of new
physics.  Studies of $W$ production at the Tevatron lead to determinations
of the mass and width of the $W$ boson with precision competitive with
LEP2 measurements~\cite{widthCDF,massCDF,widthD0,massD0}.  The ratio of
production cross sections for $W$ and $Z$ bosons, weighted by their
leptonic branching fractions, is very accurately predicted in the Standard
Model, and has been studied extensively at the Tevatron~\cite{ratioWZ}.
The rapidity distribution for produced $Z$ bosons~\cite{ZrapCDF}, 
and the charge asymmetry in leptons from $W$ production~\cite{WchargeCDF}, 
have also been measured at the Tevatron; both distributions are sensitive to 
the distribution of partons within the proton.  
Searches for non-standard contributions to the production
rate of lepton pairs with invariant masses larger than $M_{W,Z}$ can be
used to detect additional gauge bosons, such as the $Z^{'}$ states that
appear in most extensions of the SM.  More generally, these searches
constrain possible contact interactions between quarks and leptons arising
from new physics at energy scales beyond those currently accessible
\cite{ZpCDF}.

With Run II of the Tevatron producing data, and with the LHC scheduled to
begin operation shortly, an enormous number of $W$ and $Z$ bosons will
soon be collected.  This will significantly increase the precision of
electroweak measurements, and will dramatically boost the sensitivity of
new physics searches.  To fully utilize these results, precise theoretical
predictions for $W$ and $Z$ cross sections are needed.  Current
calculations are limited by uncertainties in parton distribution
functions, as well as higher-order QCD and EW radiative
corrections~\cite{WZewkcorrs}.

Parton distribution functions (PDFs) are determined from a global fit to a
variety of data; unfortunately there is no direct experimental information
for the combined values of $Q^2 \, (10^4~{\rm GeV}^2)$ and Bjorken 
$x$ ($10^{-4}$ to $10^{-1}$) that are relevant for electroweak physics at the
LHC. PDFs for these parameter values are obtained through perturbative
evolution of fits to PDFs at lower values of $Q^2$, using the DGLAP
equation.  The complete results for the DGLAP evolution kernels at
next-to-next-to leading order (NNLO) are not yet available.  An
approximate set of evolution kernels is used
instead~\cite{vanNeervenVogt}.

There are currently two sets of PDFs extracted with NNLO precision, using
these approximate kernels.  The MRST set~\cite{Martin2002} utilizes a
broad variety of data; the drawback of this procedure is that the data set
includes observables for which NNLO QCD corrections are not known.
Alekhin's PDFs~\cite{Alekhin2002} are based on deep-inelastic
scattering data only; this data set is somewhat restricted, but higher
order QCD corrections can be included consistently~\cite{NNLODIS}.  The
two PDF sets lead to slightly different (at the few percent level)
predictions for the total rate of $W$ and $Z$ production at the Tevatron
and the LHC~\cite{Martin2003,Alekhin2003}.  We take this difference
as a rough estimate of the current uncertainties in the PDFs needed for
Tevatron and LHC physics; the individual PDF fits now also contain
intrinsic uncertainty estimates~\cite{Alekhin2002,Martin2003}.

The QCD corrections to EW gauge boson production have been studied by
several groups.  The complete NNLO corrections to the total cross section
were computed some time ago~\cite{neerven,HarlanderKilgore}.  However, the
total cross section is not an experimental observable, and significant
extrapolations are required to compare this prediction to experiment.
Ideally, one would like an event generator, at least at the parton level,
which retains the full kinematics of the process and incorporates
higher-order radiative corrections.  Although there has been some recent
progress towards this goal at 
NNLO~\cite{Kosower,Weinzierla,Weinzierl,Gehrmann-DeRidder,Anastasiou2003}, 
its completion will probably not occur for some time.

NLO QCD corrections to more differential quantities in EW gauge boson
production, including the vector boson rapidity distribution, were
computed in Ref.~\cite{nlo}.  A generalization of this result to $W$ and
$Z$ production at the Tevatron and the LHC yields NLO corrections of
approximately 20\% to 50\%, and scale variations of a few percent.  Since
the NLO corrections are rather large, while the residual scale dependence
is small, the actual reliability of the NLO results has been somewhat
unclear.  Even taking the NLO scale variations seriously,
it is apparent that our knowledge of higher-order QCD corrections
to EW gauge boson production is accurate to at best a few percent.

This few percent precision in our knowledge of both PDFs and radiative
corrections must be compared to the needs of the Tevatron and LHC physics
programs.  The $W$ mass should be measured with a precision of $\pm 30$
MeV during Run II in each Tevatron experiment~\cite{Baur2003}; this
uncertainty will be further decreased to $\pm 15$ MeV at the
LHC~\cite{azuelos}.  Such a measurement strengthens the constraints that
the precision EW data imposes on the Higgs mass and on many indirect
manifestations of new physics.  A precise theoretical prediction for $M_W$
requires knowledge of the $W$ transverse momentum spectrum, as well as a
good understanding of the relevant PDFs.  To calibrate the detector
response for the measurement of the $W$ decay products, both the rapidity
and $p_\perp$ spectra of the $Z$ must also be well
understood~\cite{Baur2003}

At the LHC, many additional measurements will also require theoretical
predictions accurate to a percent or better.  The extremely large cross
section for the Drell-Yan process at the LHC allows measurements of the
$Z$ boson rapidity distribution, and of the pseudorapidity distribution 
of charged leptons originating from $W$ decays, to constrain PDFs 
at the percent level.  In effect, $W$ and $Z$ production can serve
as a parton-parton ``luminosity monitor''~\cite{plum}.  
The inferred parton-parton luminosities can then be used to precisely
predict rates for interactions with a similar initial state as
Drell-Yan production, such as gauge boson pair production processes.
Uncertainties in the overall proton-proton luminosity, which is hard
to measure precisely at the LHC, will cancel out in this approach.

It is apparent from the above examples that the Tevatron and LHC physics
programs require NNLO calculations for differential distributions in
kinematic variables; knowledge of inclusive rates is insufficient.
Although the inclusive rates for several processes, including Drell-Yan
production of lepton pairs, are known at NNLO in QCD, until very recently
no complete calculation of a differential quantity existed at NNLO.  Such
a calculation is quite challenging technically, and traditional methods
for the computation of phase-space integrals cannot handle problems of
this complexity.  In Ref.~\cite{gammastar} we described a powerful new
method of performing such calculations, and applied it to Drell-Yan
production in fixed target experiments.  We present here in detail the
computation of the rapidity distributions for Drell-Yan production of
lepton pairs through $W$ and $(Z,\gamma^*)$ exchange at both the Tevatron
and the LHC through NNLO in QCD.  Although these distributions are still
not the fully differential results needed for a Monte Carlo event
generator, they allow a large number of the physics issues discussed above
to be addressed.

Our method extends the optical theorem to allow the tools developed for
multi-loop computations to be applied to the calculation of differential
distributions.  We represent the rapidity constraint by an effective
``propagator.''  This propagator is constructed so that when the imaginary
part of the forward scattering amplitude is computed using the optical
theorem, the ``mass-shell'' constraint for the ``particle'' described by
this propagator is equivalent to the rapidity constraint in the phase
space integration.  We then use the methods described in
Ref.~\cite{htotal} for computing total cross sections, keeping the fake
particle propagator in the loop integrals, and deriving the rapidity
distribution as the imaginary part of the forward scattering amplitude.
We remark that the rapidity distribution for inclusive production 
of Higgs bosons at hadron colliders, which in the heavy top quark 
approximation is known at NLO~\cite{hrap}, can be computed at NNLO 
by precisely the same technique; indeed, all the basic integrals
encountered in the two problems are identical.

We find that the NNLO corrections to the $W$ and $Z$ rapidity
distributions are small for most values of rapidity.  This is consistent
with the results found in Ref.~\cite{neerven} for the inclusive cross
section.  However, the magnitude of the corrections can reach a few
percent for certain invariant masses and collision energies, indicating
that they are required for the precision desired in experimental analyses.
The residual scale dependences of the rapidity distributions are below the
percent level for all but the largest physically allowed rapidities.  The
theoretical uncertainty is therefore dominated by our imprecise knowledge
of the PDFs.  We study the effect of varying the PDF parameterization; we
use several fits provided by both MRST and Alekhin.  The different MRST
sets yield results for the rapidity distributions that vary by $\approx
1\%$ at the LHC; the Alekhin set gives results that differ from those of
MRST by 2--8.5\% as the rapidity is varied.  The anticipated experimental
uncertainties at the LHC are sufficiently small to distinguish between
such PDF sets.  EW gauge boson production can therefore provide important
information about the PDFs at the values of $Q^2$ and $x$ relevant for
collider experiments.  Finally, we study the efficacy of various
approximations to the complete NNLO result.  We find that the common
approximation of including only soft gluon corrections does not accurately
reproduce the full result for phenomenologically interesting parameter
choices.

Our paper is organized as follows. In Section II we introduce our
notation.  We discuss our method of calculation in detail in Section III.
We describe the collinear renormalization of the partonic cross section in
Section IV.  In Section V we present some analytic results for the
partonic rapidity distributions.  Numerical results for the $W$ and $Z$
rapidity distributions at both the Tevatron and the LHC are given in
Section VI.  We present our conclusions in Section VII.

\section{Notation}
\label{sec:notation}

We consider the production of electroweak vector bosons $V$ 
at hadron colliders,
\begin{equation}
\label{eq:reaction}
h_1 + h_2 \to V + X,
\end{equation}
where $X$ stands for any number of additional hadrons,
or partons in the perturbative calculation.
The $\bar q_i q_j V$ coupling at the tree level 
is described by the interaction vertex
\begin{equation}
\label{eq:vertex}
{\cal V}_{ij}^\mu = i g_V C_{ij} \gamma^\mu 
\left(v_i^{V} + a_i^{V} \gamma_5 \right),
\end{equation}
where the indices $i,j$ denote the quark flavors:
\begin{equation}
i,j = \{u, \bar{u}, d, \bar{d}, \ldots \}.
\end{equation}
The matrix $C_{ij}$ is the unity matrix when $V=\gamma,Z$ and 
is the CKM matrix when $V=W$. Numerical values for the 
required CKM matrix elements are given in 
Section~\ref{sec:numerics}.

The vector and axial coefficients for up and down type quarks are:
\begin{eqnarray}
&& v_{u}^{\gamma} = \frac{2}{3},~~~a_u^{\gamma} = 0,
~~~~v_{d}^{\gamma} = -\frac{1}{3},~~~a_d^{\gamma} = 0,
\nonumber \\
&& v_{u}^{Z} = 1-\frac{8}{3}\sin^2 \theta_W,~~~a_u^{Z} = -1,
~~~~v_{d}^{Z} = -1+\frac{4}{3}\sin^2 \theta_W,~~~a_d^{Z} = 1,
\nonumber \\
&&v_u^{W}=v_d^{W} = \frac{1}{\sqrt{2}},~~~~
a_u^{W}=a_d^{W} = -\frac{1}{\sqrt{2}}.
\label{vadefs}
\end{eqnarray}
The rapidity of the vector boson $V$ is defined as
\begin{equation}
\label{eq:rapidity}
Y \equiv \frac{1}{2} \log \left(\frac{E+p_z}{E-p_z} \right),
\end{equation}
where $E$ and $p_z$ are respectively the energy and longitudinal momentum 
of $V$ in the center-of-mass frame of the colliding hadrons.  
The cross section for the production of the vector boson can be written as 
the convolution of partonic hard scattering cross sections with hadronic 
parton distribution functions:
\begin{eqnarray}
\sigma^V &=& \int_{\frac{1}{2}\ln\tau}^{\frac{1}{2}\ln\frac{1}{\tau}} dY  
{d \sigma^V \over dY} \,,
\nonumber \\
 {d \sigma^V \over dY} &=&  \sum_{ab} \int_{\sqrt{\tau} e^Y}^{1}
\int_{\sqrt{\tau} e^{-Y}}^{1} dx_1 dx_2\ f_a^{(h_1)}(x_1) f_b^{(h_2)}(x_2)
\ \frac{d \sigma^V_{ab}}{dY}(x_1,x_2). 
\label{eq:xsection}
\end{eqnarray}
Here, 
\begin{equation}
\tau = \frac{m_V^2}{S},
\end{equation}
$m_V$ is the invariant mass of $V$, and $S=(P_1+P_2)^2$ is the 
square of the center-of-mass energy of the colliding hadrons 
$h_1$ and $h_2$, which carry momenta $P_1$ and $P_2$ respectively. 

As we will see in the next Section, it 
is  beneficial to represent the rapidity constraint in a 
covariant form.  To do so, we introduce the variable $u$, where
\begin{equation}
\label{eq:u_def}
u=\frac{x_1}{x_2} e^{-2 Y}.
\end{equation}
In the center-of-mass frame of the two colliding hadrons, it takes the 
simple Lorentz invariant form 
\begin{equation}
u = \frac{p_V\cdot p_1}{p_V \cdot p_2} \,,
\label{eq:u_pp_ratio}
\end{equation}
where $p_1=x_1 P_1$ and $p_2=x_2 P_2$ are the momenta of the incoming 
partons, and $p_V$ is the momentum of $V$.  The partonic center-of-mass
energy is $s = (p_1+p_2)^2 = S x_1 x_2$.  The partonic 
$u$-distributions are obtained by integrating the partonic matrix elements 
over the phase space of the final-state particles with a 
fixed value of $u$:
\begin{equation}
\label{eq:partonic_xsections}
\frac{d \sigma^V_{ab}}{d u} = \frac{1}{2s} \int d \Pi_f\ \overline{
\left\|
{\cal M}_{ab \rightarrow V+X}\right\|^2} \ \delta \left(
\frac{p_V \cdot p_1}{p_V \cdot p_2} - u
 \right).
\end{equation}
Here, $\overline {\left\|{\cal M}_{ab \rightarrow V+X}\right\|^2}$ 
denotes the square of the scattering amplitude, averaged over spins 
and colors of the colliding partons.

The allowed values of $u$ are 
\begin{equation}
\label{eq:urange}
z \leq u \leq \frac{1}{z},
\end{equation}
with 
\begin{equation}
\label{eq:z_def}
z=\frac{m_V^2}{s} = \frac{\tau}{x_1 x_2}
\end{equation}
and 
\begin{equation}
   \tau \leq z \leq 1.
\end{equation}
Inverting Eqs.~(\ref{eq:u_def}) and (\ref{eq:z_def}), the arguments
$(x_1,x_2)$ of $d \sigma_{ab}^V/dY$ in Eq.~(\ref{eq:xsection})
are given by 
\begin{equation}
\label{eq:x1x2invert}
  x_1 = { \sqrt{\tau} e^Y \over \sqrt{z/u} } \,, \qquad
  x_2 = { \sqrt{\tau} e^{-Y} \over \sqrt{uz} } \,.
\end{equation}

\begin{figure}[htbp]
\vspace{0.0cm}
\centerline{
\psfig{figure=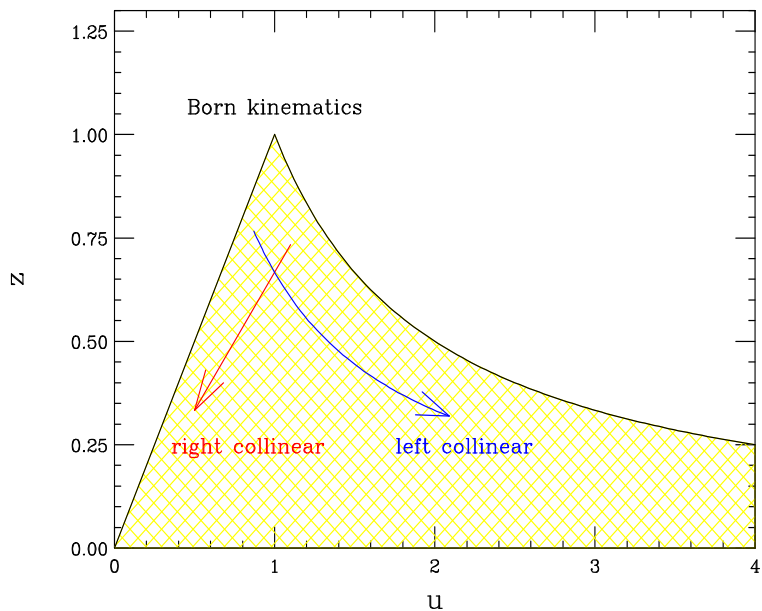,height=8.0cm,width=10.0cm,angle=0}}
\caption{Variables $(z,u)$ used to describe the kinematics of the vector
boson rapidity distribution at parton level.  The physical region is
hatched.  The point $u=z=1$ corresponds to no additional radiation, or
Born-level kinematics.  The left edge, $u=z$, corresponds to
radiation of partons collinear with incoming parton 2. The right edge, 
$u=1/z$, corresponds to radiation collinear with parton 1.
The arrows show flows relevant for the convolution integrals encountered
in mass factorization (see Section~\ref{sec:collinear}).}
\label{uzplane}
\end{figure}

The boundary values of $(z,u)$ are only achieved for special
kinematics (see Fig.~\ref{uzplane}).
For $z = 1$, $m_V^2 = s$, and there can be 
no additional partons radiated in the $V$ boson production process; 
the kinematics is that of the Born-level process $q\bar{q} \to V$.
We refer to the limit $z \to 1$ as the {\it soft} limit, since any 
additional partons must carry little energy.
The boundary $u=z$ corresponds to production of a $V$ boson along
with one or more partons radiated collinear with incoming parton 2, 
with momentum $(1-z)p_2$.  That is, inserting $p_V = p_1+zp_2$ into 
Eq.~(\ref{eq:u_pp_ratio}) leads to $u=z$.
Similarly, the boundary $u=1/z$ is achieved when the additional
partonic radiation is collinear with incoming parton 1.
We refer to the limits $u \to z$ and $u \to 1/z$ as {\it collinear} 
limits.

\noindent

\section{Method}
\label{sec:method}
We evaluate the partonic rapidity distributions of
Eq.~(\ref{eq:partonic_xsections}) through NNLO in perturbative QCD.  The
NLO corrections have been previously evaluated in Ref.~\cite{nlo}.
However, the calculation of the NNLO corrections is intractable using
current techniques.  We describe here a new method powerful enough to
handle this problem.  We express the rapidity constraint as the mass-shell
condition of a fake ``particle.''  This permits the use of the optical
theorem to transform the matrix elements into cut forward scattering
amplitudes.  We can then apply methods developed for multi-loop
integration to evaluate these amplitudes.  We describe in detail below
the required modification of the rapidity constraint, the simplification
of the forward scattering amplitude using integration-by-parts reduction
algorithms, and the evaluation of the resulting master integrals.

We begin by describing the three distinct contributions that enter at NNLO,
to illustrate the difficulties that arise:
\begin{itemize}
\item Virtual-Virtual, which contains interferences of diagrams with only
the electroweak boson $V$ in the final state;   
\begin{center}
\begin{picture}(80,35)(0,0)
\SetColor{Red}
\Line(-10,30)(35,15)
\Line(-10,0)(35,15)
\Line(70,15)(85,0)
\Line(70,15)(85,30)
\SetColor{Brown}
\Gluon(-1,3)(-1, 27){2}{4}
\Gluon(8,6)(8,24){2}{3}
\SetColor{Blue}
\SetWidth{1}
\Photon(35,15)(50, 15){2}{3}
\Photon(60,15)(70, 15){2}{3}
\end{picture}
\end{center}

\item Real-Virtual, which contains interferences with the $V$ boson and 
one additional quark or gluon in the final state;
\begin{center}
\begin{picture}(80,35)(-10,0)
\SetColor{Red}
\Line(0,5)(25,5)
\Line(0,30)(25,30)
\Line(25,5)(25,30)
\Line(65,5)(65,30)
\Line(65,5)(80,5)
\Line(65,30)(80,30)
\SetColor{Brown}
\Gluon(12.5,5)(12.5,30){2}{5}
\Gluon(40,5)(25,5){2}{3}
\Gluon(65,5)(50,5){2}{3}
\SetColor{Blue}
\SetWidth{1}
\Photon(25,30)(40,30){2}{3}
\Photon(50,30)(65,30){2}{3}
\end{picture}
\end{center}

\item Real-Real, which contains interferences of tree-type diagrams with 
$V$ and two 
additional partons in the final state. \\
\begin{center}
\begin{picture}(130,37)(5,0)
\SetColor{Red}
\Line(40,5)(60,5)
\Line(60,5)(60,35)
\Line(40,35)(60,35)
\SetColor{Blue}
\SetWidth{1}
\Photon(60,35)(80,35){2}{3}
\Photon(90,35)(110,35){2}{3}
\SetWidth{0.5}
\SetColor{Red}
\Line(110,35)(130,35) \Line(110,5)(110,35)
\Line(110,5)(130,5)
\SetColor{Brown}
\Gluon(80,20)(60,20){2}{3}
\Gluon(80,5)(60,5){2}{3}
\Gluon(110,20)(90,20){2}{3}
\Gluon(110,5)(90,5){2}{3}
\end{picture}
\end{center}

\end{itemize}

The Virtual-Virtual contribution to the rapidity distribution is identical
to its counterpart in the total cross section, which has been computed
previously~\cite{neerven}.  The new features of the rapidity distribution
are the Real-Virtual and Real-Real components, which now have non-trivial
kinematic constraints.  Until very recently no systematic technique for
their evaluation existed; this was the major reason for the lack of
progress.  However, since the calculation of the inclusive Drell-Yan cross
section, our ability to calculate diagrams of the Virtual-Virtual type has
progressed greatly.  New algorithms for the evaluation of two-loop
diagrams of the same~\cite{baikov} and more complicated
topologies~\cite{xbox,laporta} have been developed.  It is now well
understood how to organize the evaluation of generic multiloop amplitudes
using integration by parts and Lorentz invariance reduction
algorithms~\cite{laporta,ibp,tkachov,diffeq}, and how to compute the
resulting master integrals using either the
Mellin-Barnes~\cite{smirnov,tausk} or the differential equation
method~\cite{diffeqBDKK,diffeq}.
Our method renders the Real-Virtual and the Real-Real contributions 
amenable to the same techniques.

\subsection{Construction of the modified forward scattering amplitude}
\label{sec:fsa}

We follow the approach introduced in 
Ref.~\cite{htotal,atotal,hrap,gammastar}; we 
replace all non-trivial phase-space integrations by loop integrations.
To accomplish this we represent all delta functions 
constraining the final-state phase space by
\begin{equation}
\label{eq:cutkosky}
  \delta(x) 
 = \frac{1}{2\pi i} \left ( \frac{1}{x-i0} -\frac{1}{x+i0} \right ). 
\end{equation}
In the evaluation of total cross sections we only have delta functions 
which put the final-state particles on their mass shell,  
\begin{equation}
\label{eq:phasespacef}
  \int d \Pi_f \propto \prod_f \int d^d p_f\ \delta^+(p_f^2 - m_f^2), 
\end{equation}
where we work in $d=4-2\epsilon$ dimensions, 
and $\delta^+$ includes the positive energy condition $E_f > 0$.
Using Eq.~(\ref{eq:cutkosky}), each such delta function becomes a 
difference of two propagators with opposite prescription for their 
imaginary part:
\begin{equation}
\label{eq:onshell}
	\delta(p_f^2 - m_f^2)\ \to\ \frac{1}{p_f^2-m_f^2-i0} - (c.c.). 
\end{equation}
When calculating differential quantities, there are additional constraints
on the phase space.  The distribution constraints can usually be expressed
through delta functions with Lorentz-invariant arguments that are
polynomial in the momenta of the final-state particles; we then transform
them into propagators using Eq.~(\ref{eq:cutkosky}).

For the rapidity distribution of the massive boson we substitute
\begin{equation}
\label{eq:sub_u}
 \delta \left(
\frac{p_V \cdot p_1}{p_V \cdot p_2} - u
 \right)\ \to\  
\frac{p_V \cdot p_2}{p_V \cdot \left( p_1 - u p_2\right) -i0} - (c.c.).
\end{equation} 
The above substitution introduces a propagator with a scalar product in
the numerator and a denominator linear in the momentum of $V$.  However,
the multi-loop methods we employ are not sensitive to such irregularities
in the form of the propagators; they only require that the propagator of
Eq.~(\ref{eq:sub_u}) be polynomial in the momenta.  Substituting
Eqs.(\ref{eq:onshell}) and (\ref{eq:sub_u}) into
Eq.~(\ref{eq:partonic_xsections}), we obtain a forward scattering
amplitude with ``cut'' propagators originating from both the on-shell
conditions on the final-state particles and the rapidity
constraint. Pictorially, the three different contributions can be
represented by diagrams similar to the following ones:
\begin{itemize}
\item Virtual-Virtual; \\
\begin{center}
\begin{picture}(300,35)(-30,0)
\SetColor{Red}
\Line(-10,30)(35,15)
\Line(-10,0)(35,15)
\Line(120,30)(165,15)
\Line(120,0)(165,15)
\Line(70,15)(85,0)
\Line(70,15)(85,30)
\Line(190,15)(205,0)
\Line(190,15)(205,30)
\SetColor{Brown}
\Gluon(129,3)(129, 27){2}{4}
\Gluon(138,6)(138,24){2}{3}
\Gluon(-1,3)(-1, 27){2}{4}
\Gluon(8,6)(8,24){2}{3}
\SetColor{Blue}
\put(100, 15){$\Rightarrow$}
\SetWidth{1}
\Photon(35,15)(50, 15){2}{3}
\Photon(60,15)(70, 15){2}{3}
\Photon(165, 15)(177.5,15){2}{3}
\Line(177.5, 15)(190, 15)
\SetColor{Magenta}
\DashLine(177.5, 35)(177.5, 0){2}
\end{picture}
\end{center}
\item Real-Virtual; \\
\begin{center}
\begin{picture}(300,35)(-30,0)
\SetColor{Red}
\Line(0,5)(25,5)
\Line(0,30)(25,30)
\Line(25,5)(25,30)
\Line(65,5)(65,30)
\Line(65,5)(80,5)
\Line(65,30)(80,30)
\Line(120,5)(155,5)
\Line(185,5)(200,5)
\Line(155,5)(155,30)
\Line(185,5)(185,30)
\Line(120,30)(155,30)
\Line(185,30)(200,30)
\SetColor{Brown}
\Gluon(12.5,5)(12.5,30){2}{5}
\Gluon(40,5)(25,5){2}{3}
\Gluon(65,5)(50,5){2}{3}
\Gluon(185,5)(155,5){2}{6}
\Gluon(132.5,5)(132.5,30){2}{5}
\SetColor{Blue}
\put(100,15){$\Rightarrow$}
\SetWidth{1}
\Photon(25,30)(40,30){2}{3}
\Photon(50,30)(65,30){2}{3}
\Photon(155,30)(170,30){2}{3}
\Line(170,30)(185,30)
\SetColor{Magenta}
\DashLine(170,35)(170,0){2}
\end{picture}
\end{center}
\item Real-Real.\\
\begin{center}
\begin{picture}(300,35)(-30,0)
\SetColor{Red}
\Line(10,5)(25,5)
\Line(10,30)(25,30)
\Line(25,5)(25,30)
\Line(65,5)(65,30)
\Line(65,5)(80,5)
\Line(65,30)(80,30)
\Line(140,5)(155,5)
\Line(185,5)(200,5)
\Line(155,5)(155,28)
\Line(185,5)(185,28)
\Line(140,30)(155,30)
\Line(185,30)(200,30)
\SetColor{Brown}
\Gluon(40,5)(25,5){2}{3}
\Gluon(40,18)(25,18){2}{3}
\Gluon(65,18)(50,18){2}{3}
\Gluon(65,5)(50,5){2}{3}
\Gluon(185,5)(155,5){2}{6}
\Gluon(185,18)(155,18){2}{6}
\SetColor{Blue}
\put(100,15){$\Rightarrow$}
\SetWidth{1}
\Photon(25,30)(40,30){2}{3}
\Photon(50,30)(65,30){2}{3}
\Photon(155,30)(170,30){2}{2}
\Line(170,30)(185,30)
\SetColor{Magenta}
\DashLine(170,35)(170,0){2}
\end{picture}
\end{center}
\end{itemize}
We have associated an additional ``rapidity'' propagator with the $V$
boson, which we represent by a straight line just to the right of
the cut from the usual wavy (cut) $V$ boson propagator. 
In the above diagrams, cut propagators represent differences of two
complex conjugate terms, propagators on different sides of the cut have
different prescriptions for their imaginary part, and the initial and final
states are identical.

The three contributions are now expressed as two-loop amplitudes in which
the cuts denote differences of propagators with opposite $i\epsilon$
prescriptions.  These cut conditions are accounted for at the very end of
the calculation, after using generic multi-loop methods to simplify the
two-loop expressions.  We generate the diagrams for the forward scattering
amplitude using QGRAF~\cite{qgraf}. We then apply the Feynman rules,
introduce the rapidity ``propagator'' of Eq.~(\ref{eq:sub_u}), and perform
color and Dirac algebra (we use conventional dimensional regularization)
using FORM~\cite{form}.  This generates a large number of integrals
with cut propagators which we must evaluate.  The evaluation of these
integrals is discussed in the next Section.  Our treatment of $\gamma_5$
in dimensional regularization follows the discussion in
Ref.~\cite{neerven}, to which we refer the reader for further details.

\subsection{Reduction to Master Integrals}
\label{sec:reduction}

An essential part of the calculation is the reduction of the
integrals to a small set of independent master integrals using linear
algebraic relations among them.  This procedure is routinely applied in
computations of virtual amplitudes, such as in the Virtual-Virtual
contribution. Our representation of the Real-Virtual and Real-Real terms
makes it possible to evaluate them in a similar manner.
 
We first apply partial fractioning identities to reduce the denominators
of the integrands to a linearly independent set.  Partial fractioning is
always applicable to box diagrams with the introduction of the rapidity
constraint.  Seven independent scalar products can be formed from the two
external momenta $p_1,p_2$ and the two loop momenta $k,l$ (omitting the
constant scalar products, $p_i\cdot p_j$).  On the other hand, box
topologies with the rapidity propagator have eight terms in the
denominator.  Thus the eight terms obey one linear relation, which can be
used to perform a partial fraction decomposition whenever all terms
entering the relation appear as denominators.  This step allows us to
eliminate one of the uncut propagators in favor of the rapidity
propagator. (Whenever a cut propagator or the rapidity propagator does not
appear in the denominator of an integral, that integral may be set to
zero, by anticipating the delta function constraints.)  From this
procedure we derive 11 major topologies for the Real-Real contributions, 2
major topologies for the Real-Virtual contributions, and 2 for the
Virtual-Virtual contributions. All non-box diagrams are sub-topologies of
the above set.

We obtain additional recurrence relations using integration-by-parts (IBP)
identities~\cite{ibp,tkachov}. If $k,l$ are the loop-momenta of the
two-loop integrals in the forward scattering amplitude, and $p_1,p_2$ are
the incoming momenta, we can write 8 IBP identities of the following form
for each integral:
\begin{equation}
\label{eq:ibp}
0=\int d^dk\ d^dl\ \frac{\partial}{\partial \eta_\mu}
\ \frac{q^\mu}{k^2 l^2 \ldots} \,,
\end{equation}
with $\eta^\mu = k^\mu, l^\mu$ and $q^\mu = k^\mu, l^\mu, p_1^\mu, p_2^\mu$.  
Each integral contains some cut propagators. However,
since differentiation with respect to the loop momenta is insensitive to
the prescription for the imaginary part of propagators, the application of
IBP reduction algorithms and the taking of cuts
commute~\cite{htotal}. This fact allows a straightforward derivation of
reduction relations for phase-space integrals.  Similarly, we can derive
Lorentz invariance identities~\cite{diffeq} for phase-space integrals;
however, for the Drell-Yan rapidity distribution they are not linearly 
independent of the IBP relations and provide no additional information.

To solve the system of equations formed by the IBP relations we use an
algorithm introduced by Laporta~\cite{laporta}.  We construct a large
system of explicit IBP identities which we then solve using Gauss
elimination.  This system should include a sufficient number of equations
to reduce all the integrals of the forward scattering amplitudes to master
integrals.  A detailed description of the algorithm can be found in the
original paper of Laporta~\cite{laporta}; we have implemented a customized
version in MAPLE~\cite{maple} and FORM~\cite{form}.  An important
simplification of the reduction procedure in the present case is that we
can discard all integrals in which the cut propagators, or the rapidity
propagator, are eliminated or appear with negative powers~\cite{htotal}.
After performing the reductions, we obtain 5 master integrals for the
Virtual-Virtual contributions, 5 for the Real-Virtual, and 19 for the
Real-Real.

\subsection{Master integrals}
\label{sec:masters}

All Virtual-Virtual master integrals were known prior to this calculation.  
The only non-trivial one is the crossed-triangle master integral, which was 
calculated in Ref.~\cite{gonzalves,kramer}. A list of all Virtual-Virtual 
master integrals can be found in the appendix of ~\cite{htotal}. 

For the Real-Virtual contributions we find the following master integrals:
\[
\YEb{0} \YBb{0} \YAb{0}  \YCb{0} \YDb{0}  
\]
where solid lines correspond to massless scalar propagators 
\[
\begin{picture}(30,30)(0,8)
\Line(5,10)(28, 10)
\end{picture}
\quad
\to \quad \frac{1}{p^2},
\] 
bold solid lines correspond to massive scalar propagators 
\[
\begin{picture}(30,30)(0,8)
\SetWidth{2}
\Line(5,10)(28, 10)
\end{picture}
\quad
\to \quad \frac{1}{p^2 - m_V^2},
\]
and dashed lines denote the rapidity propagator
\[
\begin{picture}(30,30)(0,8)
\DashLine(5,10)(28, 10){1.5}
\end{picture}
\quad
\to \quad \frac{1}{p_V \cdot \left[p_1 - u p_2 \right]}.
\]
The Real-Virtual master integrals can be evaluated by using
Eq.~(\ref{eq:cutkosky}) to reinstate the delta-function constraints.  We
then must perform a one-loop integral and a 2-particle phase-space
integral; both are straightforward.  The most complicated loop integral is
a massless one-loop box diagram with one external leg off-shell, which is
known to all orders in $\epsilon$~\cite{ndimbox}.  The Real-Virtual
phase-space integration is simple because the polar angle for the 
2 $\to$ 2 process is fixed by the rapidity constraint, leaving only a
$(1-2\epsilon)$-dimensional azimuthal angular integration.  It
is thus straightforward to derive analytic expressions for the
Real-Virtual master integrals which are valid to all orders in $\epsilon$.

The Real-Real phase-space master integrals were unknown prior to this
calculation.  A few can be evaluated directly; for example, the two
simplest master integrals,
\begin{eqnarray}
&& I[0] =  \Xone{30} = \int d^d q_V d^dq_1 d^d q_2
\delta^d(p_1+p_2-q_V-q_1-q_2) \times
\nonumber \\
&& \nonumber \\
&& \nonumber \\
&& \hspace{2cm}	 
  \delta^+(q_V^2-m_V^2) \delta^+(q_1^2) \delta^+(q_2^2)
	\delta(q_V \cdot\left[p_1-u p_2 \right] )  
\end{eqnarray}
and 
\begin{eqnarray}
&& I[1] =  \Xfourteen{30} = \int d^d q_V d^dq_1 d^d q_2 
\delta^d(p_1+p_2-q_V-q_1-q_2) \times
\nonumber \\
&& \nonumber \\
&& \nonumber \\
&& \hspace{2cm}	 
	(q_1+q_2)^2 
  \delta^+(q_V^2-m_V^2) \delta^+(q_1^2) \delta^+(q_2^2)
	\delta(q_V \cdot\left[p_1-u p_2 \right] )  
\end{eqnarray}
have the following hypergeometric integral representation: 
\begin{equation}
\label{eq:basicmasters}
I[\nu] = \frac{ \Omega_{d-2} \Omega_{d-1} s^{\nu-2\epsilon}}
{2^d (1+u)^{1-2\epsilon}}
\frac{(u-z)^{1+\nu-2\epsilon} (1-uz)^{1+\nu-2\epsilon} }
{\left[ \sqrt{u} (\sqrt{u}+\sqrt{z}) (1+\sqrt{uz})\right]^{1+\nu-\epsilon}} 
{\cal K}_\nu(\delta),
\end{equation}
with 
\begin{equation}
\Omega_d = { 2 \pi^{d/2} \over \Gamma(d/2) } \,,
\end{equation}
\begin{equation}
{\cal K}_{\nu}(\delta) = \int_0^1 d\chi \chi^{\nu} \left[ 
\chi (1-\chi) (1-\chi \delta)\right]^{-\epsilon} \,,
\end{equation}
and 
\begin{equation}
\delta = \frac{(\sqrt{u}-\sqrt{z}) (1-\sqrt{uz})}
{(\sqrt{u}+\sqrt{z}) (1+\sqrt{uz})}.
\end{equation} 
Expanding in $\epsilon$, we obtain
\begin{eqnarray}
{\cal K}_0(\delta) &=& 1+\epsilon\, 
\left[ 3 - {\frac {\left( \delta-1 \right) 
                \ln \left( 1-\delta \right) }{\delta}} \right] 
+{\epsilon}^{2} \Bigg[ {\frac {
 \left( \delta-1 \right)  \ln^2  \left( 1-\delta \right) }{\delta}}
\nonumber \\
&& -3\,{\frac {\left( \delta-1 \right) \ln \left( 1-\delta \right) 
  }{\delta}}+{\frac { \left( \delta-2 \right) {
\rm Li}_2 \left(\delta \right) }{\delta}}+9-\frac{{\pi }^{2}}{6}
 \Bigg] + {\cal O} (\epsilon^3),
\end{eqnarray}
and 
\begin{eqnarray}
{\cal K}_1(\delta) &=& \frac{1}{2}+\frac{\epsilon}{2} \, 
\left[ -{\frac { \left( \delta^2-1 \right) 
 \ln  \left( 1-\delta \right) }{{\delta}^{2}}}+{\frac {5
\,\delta+2}{2 \delta}} \right] 
+\frac{{\epsilon}^{2}}{2} \Bigg[ {\frac { \left( 
\delta^2-1 \right)    \ln^2  \left( 1-\delta
 \right) }{{\delta}^{2}}}
\nonumber \\ 
&& 
\hspace{-0.7cm}
-5\,{\frac { \left( \delta^2-1
 \right)   \ln  \left( 1-\delta \right) }{2 {\delta}^{2}}} 
+{\frac { \left( {\delta}^{2}-2 \right) {
\rm Li}_2 \left(\delta \right) }{{\delta}^{2}}}
+9\,{\frac {3\,\delta+2}{4 \delta}} 
- \frac{{\pi }^{2}}{6}
\Bigg]
  + {\cal O} (\epsilon^3).
\end{eqnarray}

In the expression for the scattering amplitude, some of the master
integrals are multiplied by coefficients which become singular at the
phase-space boundaries.  For example, when the $I[\nu]$ master integrals
get multiplied by ${1}/{(u-z)^{2+\nu}}$ or $1/{(1-uz)^{2+\nu}}$, the
matrix elements become singular at $u=z$ or $u=1/z$ respectively. These
singularities are regulated by the non-integer powers of the $(u-z)$ and
$(1-uz)$ prefactors in Eq.~(\ref{eq:basicmasters}).  Upon integrating over
$u$ and $z$, they generate $1/\epsilon$ poles which cancel against the
Real-Virtual and Real-Real $1/ \epsilon$ singularities.  For example,
\begin{equation}
\int_z^{1/z} du\ (u-z)^{-1-2\epsilon}= -\frac{1}{2\epsilon} 
\left( \frac{1-z^2}{z} \right)^{-2\epsilon} \,.
\end{equation}
Since we are interested in the rapidity distribution, we do not integrate
over $u$.  We must therefore extract these singularities from the
Real-Real master integrals.  To do so, we factor out the leading behavior
of the integral ${\cal X}_i$ in the limits $u \to z$ and $u \to 1/z$,
keeping the exact $\epsilon$-dependence:
\begin{equation}
{\cal X}_i(z,u) = (u-z)^{m-\alpha \epsilon} (1-uz)^{n - \beta \epsilon}
{\cal F}_i(z,u).
\end{equation}
The integers $m,n$ are characteristic to each master integral, while 
$\alpha=\beta=2$ for all Real-Real phase-space integrals.  The functions 
${\cal F}_i$ are smooth and non-zero at $u=z$ and $u=1/z$, and can be 
calculated as a series in $\epsilon$. In the non-singular regions of 
phase space we need only calculate the first few terms in the $\epsilon$ 
expansion, up to the order where polylogarithms of rank 2 appear. However, 
at $u=z$ and $u=1/z$ additional $1/\epsilon$ coefficients may be generated, 
and at $u=z=1$ additional $1/\epsilon^2$ poles may appear.  These require an 
$\epsilon$ expansion of ${\cal F}_i$ up to a transcendentality
of rank 3 or 4.  
We therefore split the master integrals into four different terms:
\begin{equation}
\label{eq:mastersplit}
{\cal X}_i = {\cal X}_i^{soft} 
+{\cal X}_i^{coll(z)}+{\cal X}_i^{coll(\frac{1}{z})}+{\cal X}_i^{hard}.
\end{equation}
Here, 
\begin{equation}
{\cal X}_i^{soft} =(u-z)^{m-\alpha \epsilon} (1-uz)^{n - \beta \epsilon}
{\cal F}_i(1, 1)
\end{equation}
is potentially singular at the limits  $u=z$, $u=1/z$ and $u=z=1$; 
\begin{equation}
{\cal X}_i^{coll(z)} =(u-z)^{m-\alpha \epsilon} 
(1-z^2)^{n - \beta \epsilon}
\left[ {\cal F}_i(z, z)- {\cal F}_i(1, 1) \right] 
\end{equation}
can only become singular at $u=z$; 
\begin{equation}
{\cal X}_i^{coll(\frac{1}{z})} = 
\left(\frac{1}{z} - z\right)^{m-\alpha \epsilon} 
(1-uz)^{n - \beta \epsilon}
\left[ {\cal F}_i(z, 1/z)- {\cal F}_i(1, 1) \right] 
\end{equation}
can only become singular at $u=1/z$; and 
\begin{equation}
{\cal X}_i^{hard}={\cal X}_i -{\cal X}_i^{soft} 
-{\cal X}_i^{coll(z)}-{\cal X}_i^{coll(\frac{1}{z})}
\end{equation}
is smooth in all singular limits. We extract the explicit 
$1/\epsilon$ terms from the $coll$ and $soft$ terms by replacing 
the $u$ variable with 
\begin{equation}
\label{eq:ydef}
y=\frac{u-z}{(1-z)(1+u)},
\end{equation}
where $0 \leq y \leq1$.  We then apply identities of the form 
\begin{equation}
\label{eq:plusexpansion}
x^{-1+\epsilon} = \frac{1}{\epsilon} \delta(x) + \sum_n 
\frac{\epsilon^n}{n!} \left[\frac{\ln^n x}{x} \right]_+,
\end{equation}
for $x=y,1-y$ and $1-z$.  The advantage of using  
the variable $y$ instead of  $u$ is that 
$y$ separates the singularities at $u=z$ and $u=1/z$, which 
overlap when $z=1$.   [At next-to-leading-order, and for
the Real-Virtual 2-particle phase space at NNLO, the variable $y$ is
related to the $2 \to 2$ partonic center-of-mass scattering angle
$\theta^*$ by $y=(1+\cos\theta^*)/2$.]

Although a deeper expansion in $\epsilon$ is required for the master 
integrals in the collinear and soft regions, the calculation is simplified 
since in the collinear regions the result has a non-trivial dependence
on only the variable $z$; in the soft region, the ${\cal F}_i$ have no 
dependence on either $u$ or $z$.  
For example, while it is difficult to expand ${\cal K}_\nu(\delta)$ to higher 
orders in $\epsilon$ for generic $\delta$, in the soft
region, $\delta \to 0$, it can be computed in terms of Gamma functions 
in closed form:
\begin{equation} 
{\cal K}_\nu(0)= \frac{\Gamma(1+\nu-\epsilon)\Gamma(1-\epsilon)}
{\Gamma(2+\nu-2\epsilon)}.
\end{equation}

\subsection{The differential equation method}
\label{sec:diffeq}

The two Real-Real master integrals of the previous subsection 
were calculated by deriving a simple hypergeometric integral 
representation starting from their definition as phase-space integrals. 
However, this is not practical for most master integrals.  In more complicated 
cases we resort to the method of differential equations. This method was 
developed for loop integrals~\cite{diffeqBDKK,diffeq}; 
however the representation of 
Eq.~(\ref{eq:cutkosky}) for delta function constraints allows its 
application to phase-space integrations in a straightforward 
manner~\cite{htotal}.
We consider the following master integral as an example:
\begin{eqnarray}
\label{eq:exmi}
{\cal J}(z, u) &=&  \int d^d q_V d^d q_1 d^d q_2\  
\delta^d (p_1+p_2-q_V-q_1-q_2)
\times \nonumber \\
&& \delta^+(q_V^2-m_V^2) \delta^+(q_1^2) \delta^+(q_2^2) 
\delta(q_V\cdot [p_1 - u p_2])
\frac{1}{(q_1+q_V-p_1)^2}.
\end{eqnarray}
After applying the transformation of Eq.~(\ref{eq:cutkosky}), this  
integral becomes
\begin{eqnarray}
\label{eq:exmiC}
{\cal J}(z, u) &=&\int d^dk\ d^dl\ 
\left[ \frac{1}{k^2 - m_V^2} \right]_c
\left[ \frac{1}{(k-l)^2} \right]_c  
\times \nonumber \\ 
&& \hspace{-0.4cm}
\left[ \frac{1}{(l+p_1+p_2)^2} \right]_c  
\left[ \frac{-1}{k\cdot (p_1-up_2)} \right]_c 
\frac{1}{(l+p_1)^2}, 
\end{eqnarray}
where $k = - q_V$, $l = - q_V - q_1$, and we denote 
\begin{equation}
\left[\frac{1}{x} \right]_c = \frac{1}{2 \pi i} \left( 
\frac{1}{x-i0} - \frac{1}{x+i0}
\right).
\end{equation}
We can now differentiate ${\cal J}(z, u)$ with respect to $z$ and $u$, 
obtaining 
\begin{eqnarray}
\label{eq:zinteg}
\frac {\partial {\cal J}(z, u)}{\partial z} &=& 
\int d^dk\ d^dl 
\left[ \frac{1}{\left(k^2 - m_V^2\right)^2} \right]_c
\left[ \frac{1}{(k-l)^2} \right]_c  
\times \nonumber \\ 
&& \hspace{-0.4cm}
\left[ \frac{1}{(l+p_1+p_2)^2} \right]_c  
\left[ \frac{-1}{k\cdot (p_1-up_2)} \right]_c 
\frac{1}{(l+p_1)^2}, 
\\
\label{eq:uinteg}
\frac {\partial {\cal J}(z, u)}{\partial u} &=&
\int d^dk\ d^dl 
\left[ \frac{1}{k^2 - m_V^2} \right]_c
\left[ \frac{1}{(k-l)^2} \right]_c  
\times \nonumber \\ 
&& \hspace{-0.4cm}
\left[ \frac{1}{(l+p_1+p_2)^2} \right]_c  
\left[ \frac{-1}{\left(k\cdot (p_1-up_2)\right)^2} \right]_c 
\frac{k\cdot p_2}{(l+p_1)^2}.
\end{eqnarray}
We have set $s=\left(p_1+p_2\right)^2=1$ in these expressions.  
Neither integral on the right-hand side of 
Eqs.~(\ref{eq:zinteg}) and (\ref{eq:uinteg}) is a master integral. 
However, using IBP we can reduce them to the master integrals ${\cal J}$, 
$I[0]$, and $I[1]$, using the reduction algorithm of 
Section~\ref{sec:reduction}. 
We then obtain a system of two partial differential equations which 
determines the functional dependence of ${\cal J}$ on the two kinematic 
variables $z,u$:  
\begin{eqnarray}
\label{eq:diffz}
\frac{\partial {\cal J}(z, u)}{\partial z} &=& 
\frac{2\epsilon}{u-z} {\cal}J(z, u) + 
\frac{(1-2\epsilon)u 
\left[ 
(1+3zu+4z)\epsilon -1-zu-2z
\right] }
{2 \epsilon z (u-z) (1-uz)}
I[0](z,u) \nonumber \\
&& + \frac{(1-2\epsilon)(2-3\epsilon)u}{2\epsilon z(u-z)(1-uz)} I[1](z,u),
\\
\label{eq:diffu}
\frac{\partial {\cal J}(z, u)}{\partial u} &=&
-\frac{2\epsilon}{u-z} {\cal}J(z, u) +
\frac{(1-2\epsilon) \left[
(7+4z-3zu)\epsilon -3-2z+zu
\right]
}{2\epsilon(u-z)(1-uz)} I[0](z,u)
\nonumber \\
&&
+ \frac{(1-2\epsilon)(2-3\epsilon) }
{2\epsilon (u-z)(1-uz)} I[1](z,u).
\end{eqnarray}
The general solution of Eq.~(\ref{eq:diffz}) is
\begin{equation}
\label{eq:solz}
{\cal J}(z, u) = \left[ { \Omega_{d-1} \over 2^{d-2} } \right]^2
(u-z)^{-2\epsilon}\left\{ \int^z   
dz_1
\left(u-z_1 \right)^{2\epsilon} \beta(z_1, u) +
f(u)+ {\cal C} \right\},
\end{equation}
where $(\Omega_{d-1}/2^{d-2})^2 \, \beta(z,u)$ 
is the inhomogeneous part of the differential equation in 
Eq.~(\ref{eq:diffz}). We can evaluate the integral in Eq.~(\ref{eq:solz}) 
as a series
 in $\epsilon$ after we rewrite $\beta$ using the expressions for 
$I[\nu]$ from subsection~\ref{sec:masters}.  We obtain
\begin{eqnarray}
{\cal R} &=& \int^z    
dz_1 
\left(u- z_1 \right)^{2\epsilon} \beta(z_1, u)= 
\frac{A_1(z,u)}{\epsilon} + A_0(z,u) + {\cal O}(\epsilon),
\end{eqnarray}
with 
\begin{equation}
A_1(z, u) = \frac{1}{2}\left[ \ln(r) - \ln(r+t) \right]
\end{equation}
and 
\begin{eqnarray}
\label{eq:A0zu}
A_0(z, u) &=&
-\frac{1}{2} \left[ \ln(r) + \ln(2) 
 \right] \ln ( 1+{r}^{2}) 
+\frac{1}{2} \ln ( r ) 
 \left[ 2 \ln( r) +\ln( 2) -4 \right] 
\nonumber \\ 
&&
-\frac{1}{2} \ln( t) \ln  ( r ) 
+ \frac{1}{2} \left[ \ln( 2 ) +4-\ln( r ) +\ln( {r}^
{2}+1)  \right] \ln  ( r+t ) -\frac{1}{2} \ln^2 ( r+t)
\nonumber \\
&&
+\frac{1}{2} \ln( t) \ln( r+t) 
-{\rm Li_2} \left[ {\frac { ( r+t) r}{{r}^{2}+1}} \right] 
+\frac{1}{2}  {\rm Li_2} \left[ -{\frac {t}{r}
} \right] 
+\frac{1}{2} {
\rm Li_2} \left[ {\frac {r-t}{r}} \right] 
\nonumber \\
&&
+{\rm Li_2} \left[ {\frac {2
{r}^{2}}{{r}^{2}+1}} \right] 
-\frac{1}{2}{\rm Li_2} \left[ {\frac {r-t}{2 r}} \right] 
-\frac{1}{2} {\rm Li_2} 
 \left[ {\frac { \left( r-t \right) r}{{r}^{2}+1}} \right].
\end{eqnarray}
We have introduced the notation $r=\sqrt{u}$ and $t=\sqrt{z}$.  
Substituting the solution of Eq.~(\ref{eq:solz}) 
into the differential equation 
of Eq.~(\ref{eq:diffu}), 
we derive a differential equation for $f(u)$ which we can 
again solve order by order in $\epsilon$.  We find 
\begin{equation}
f(u)= \frac{f_1(u)}{\epsilon} + f_0(u) + {\cal O}(\epsilon),
\end{equation}
with 
\begin{equation}
f_1(u)= \frac{1}{2} \ln(1+r^2) - \ln(r)
\end{equation}
and
\begin{eqnarray}
\label{eq:f0u}
f_0(u) &=&
{\rm Li_2} \left[ -{r}^{2} \right] +{\rm Li_2} \left[ \frac{{r}^{2}+1}{2}
 \right] -{\rm Li_2} \left[ {r}^{2} \right] 
+\frac{1}{4} \ln^2 ( {r}^{2}+1)- \ln^2( r )
+4\ln( r) 
\nonumber \\
&&
-\frac{1}{2} \left[\ln( 
{r}^{2}+1) +2 \ln( r)  \right] \ln( 2) +
 \left[ \ln( r ) -2 \right] \ln ( {r}^{2}+1). 
\end{eqnarray}
Finally, we must determine the constant of integration ${\cal C}$. 
In principle, this requires an explicit calculation at a specific kinematic 
point ${\cal J}(z_0, u_0)$. However, in many cases we can extract the 
constant of integration by comparing to the asymptotic behavior of all 
rapidity phase-space integrals at $u=z=1$, which is identical to that of 
the basic master integral $I[0]$:
\begin{equation}
\label{eq:psscaling}
\lim_{z,u \to 1} {\rm PS}(z,u) = c (u-z)^{n-2\epsilon}  (1-uz)^{m-2\epsilon}. 
\end{equation} 
The $\epsilon$ power of the $u-z$ and $1-uz$ factors is determined by the
number of dimensions, $d=4-2\epsilon$; adding more propagators to the
basic master integral $I[0]$ can only alter the integers $n,m$ of the
asymptotic scaling.  We note that the presence of the constant of
integration ${\cal C}$ in Eq.~(\ref{eq:solz}) violates the scaling of
Eq.~(\ref{eq:psscaling}).  We can therefore evaluate ${\cal C}$ by
requiring that all the terms in Eq.~(\ref{eq:solz}) that violate
Eq.~(\ref{eq:psscaling}) in the limit $z\to 1,u \to 1$ cancel.  We obtain
\begin{equation}
{\cal C} =\frac{1}{4} \left[\ln^22 + \frac{\pi^2}{2}  \right] 
+ {\cal O}(\epsilon).
\end{equation}  
There are master integrals for which the solution of the homogeneous
differential equation gives a scaling at $u=z=1$ which is consistent with
Eq.~(\ref{eq:psscaling}) for arbitrary values of the constant ${\cal C}$.
For these master integrals, we must determine ${\cal C}$ by performing an
explicit evaluation in the vicinity of this kinematic point.

As discussed previously, we often need to calculate master integrals in
their soft or collinear limits to higher orders in $\epsilon$.  For
example, the integral ${\cal J}$ is typically divided by an explicit
$(u-z)$ factor in the matrix elements, requiring an $\epsilon$-expansion
in its collinear limit $u \to z$ which includes the order $\epsilon$ term.
We could extend the outlined calculation of ${\cal J}$ for generic $z,u$
to include the ${\cal O}(\epsilon)$ term and then take the limit $u \to z$.  
However, this would involve expressing the result for generic $z,u$
through generalized polylogarithms of rank 3 with two variables; taking
the $u\to z$ limit would collapse them to rank 3 polylogarithms with only
the argument $z$.  We can avoid the two-variable rank 3 polylogarithms by
solving the differential equations directly in the $u \to z$ limit.  We
express the $z$-dependent term in the general solution of
Eq.~(\ref{eq:solz}) in the form
\begin{equation}
{\cal R}=-\int_z^u dz_1 (u-z_1)^{2\epsilon} \beta(z_1 , u), 
\label{quiks}
\end{equation}
and perform the change of variables 
\begin{equation}
z_1 = z + (u-z) \lambda.
\end{equation}
Next we expand the integrand in $u-z$ and keep only the leading term.
Only the $coll(z)$ limits of the boundary integrals $I[\nu]$ are required,
and as explained above those are known to all orders in $\epsilon$.  We
can then expand Eq.~(\ref{quiks}) in $\epsilon$; the resulting integration
over $\lambda$ involves polylogarithms with a single argument $z$, and can
be performed straightforwardly.  The computation of $\lim_{u\to z} f(u)$
proceeds as before, utilizing equivalent expansions in $u-z$.  Finally,
the constant ${\cal C}$ is determined by matching to the asymptotic
behavior, $(u-z)^{-2\epsilon} (1-z^2)^{-2\epsilon}$.

An important check of our results for the master integrals is provided by
integrating them over the rapidity variable $u$.  The master integrals 
also enter the NNLO corrections to the rapidity distribution for 
Higgs boson production at hadron colliders via gluon-gluon fusion,
computed in the heavy top quark approximation.
Hence the integrated master integrals can be expressed in terms of 
the master integrals appearing in the evaluation of the Higgs boson 
total cross section.  We have verified that all rapidity-distribution 
master integrals are consistent with the results of Ref.~\cite{htotal}.  
The analytic expressions for the master integrals are too lengthy to 
present here. They can be obtained from the authors by request.  

\section{Renormalization and mass factorization}
\label{sec:collinear}

The partonic cross sections of Eq.~(\ref{eq:xsection}), after combining
the real and virtual contributions up to ${\cal O}(\alpha_s^2)$, contain
$1/\epsilon^2$ and $ 1/\epsilon$ poles arising from both ultraviolet and
initial-state collinear singularities. We remove the UV singularities
through renormalization in the $\overline{{\rm MS}}$ scheme, and absorb
the initial-state singularities into the PDFs using the $\overline{{\rm
MS}}$ factorization scheme.  First we expand the cross section in the 
strong coupling constant,
\begin{equation}
\label{eq:bareexpansion}
\frac{d \sigma_{ij}}{dY} = \frac{d \hat\sigma_{ij}^{(0)}}{dY} 
+ \left ( \frac{\alpha_s^\prime}{\pi} \right )
\frac{d \hat\sigma_{ij}^{(1)}}{dY}
+ \left( \frac{\alpha_s^\prime}{\pi}\right)^2
\frac{d \hat\sigma_{ij}^{(2)}}{dY} + {\cal O}((\alpha_s^\prime)^3).
\end{equation}
The bare strong coupling $\alpha_s^\prime$ is related to the 
running strong coupling constant $\alpha_s = \alpha_s(\mu)$ in the 
$\overline{{\rm MS}}$ scheme via
\begin{equation}
\label{eq:alphasren}
\alpha_s^\prime \left(4 \pi \right)^\epsilon e ^{-\epsilon \gamma} =
\alpha_s\ \mu^\epsilon\
\left[1 -\frac{\alpha_s}{\pi} \frac{\beta_0}{\epsilon} 
+ {\cal O}(\alpha_s^2) \right], 
\end{equation}
with 
\begin{equation}
\beta_0 = \frac{11}{4} - \frac{1}{6} n_f.
\end{equation}
Here $n_f$ is the number of light quark flavors, and
$\mu = \mu_R = \mu_F$ is the combined renormalization and
factorization scale.   At the end of the calculation, we restore the 
dependence on $\mu_R$ alone, with the aid of the renormalization 
group equation.   Substituting Eq.~(\ref{eq:alphasren}) into
Eq.~(\ref{eq:bareexpansion}) and collecting with respect to 
$\alpha_s$, gives the coefficients of the renormalized expansion,
\begin{equation}
\label{eq:reneexpansion}
\frac{d \sigma_{ij}}{dY} = \frac{d \sigma_{ij}^{(0)}}{dY} 
+ \left ( \frac{\alpha_s}{\pi} \right )
\frac{d \sigma_{ij}^{(1)}}{dY}
+ \left( \frac{\alpha_s}{\pi}\right)^2
\frac{d \sigma_{ij}^{(2)}}{dY} + {\cal O}((\alpha_s)^3),
\end{equation}
in terms of the bare ones in Eq.~(\ref{eq:bareexpansion}).

Similarly, to remove the initial-state singularities,
we rewrite the hadronic cross section of Eq.~(\ref{eq:xsection}) using 
infrared-finite partonic cross sections: 
\begin{equation}
\label{eq:mfxsection}
\frac{d \sigma^V}{d Y} = 
\sum_{ab} \int_{\sqrt{\tau} e^Y}^{1}
\int_{\sqrt{\tau} e^{-Y}}^{1} dx_1 dx_2 \tilde{f}_a^{(h_1)}(x_1) 
\tilde{f}_b^{(h_2)}(x_2)
\frac{d \tilde{\sigma}^V_{ab}}{dY} (x_1,x_2).
\end{equation} 
The renormalized parton distribution functions $\tilde{f}^{(h)}_a$ 
are related to the ``bare'' ones ${f}^{(h)}_b$ by 
\begin{equation}
\label{eq:msfac}
\tilde{f}^{(h)}_a = {f}^{(h)}_b \otimes \Gamma_{ab}.
\end{equation}
We have introduced the convolution integral
\begin{equation}
\left(f \otimes g \right)(x) = \int_0^1 dy dz\  f(y) g(z) \delta(x-yz),
\end{equation}
and we implicitly sum over repeated parton indices.
The functions $\Gamma_{ab}$ are given in the $\overline{{\rm MS}}$ scheme
by
\begin{eqnarray}
\Gamma_{ab}(x)& =& \delta_{ab} \delta(1-x)
-\frac{\alpha_s}{\pi} \frac{P_{ab}^{(0)}(x)}{\epsilon} \nonumber \\
&& 
\hspace{-1cm}
+ \left( \frac{\alpha_s}{\pi} \right)^2
\left\{
\frac{1}{2 \epsilon^2} \left[ 
\left(P_{ac}^{(0)} \otimes P_{cb}^{(0)} \right)(x) + \beta_0 P_{ab}^{(0)}(x)
\right]
-\frac{1}{2\epsilon} P_{ab}^{(1)}(x)
\right\} + {\cal O}(\alpha_s^3),
\end{eqnarray}
where the Altarelli-Parisi kernels $P_{ab}^{(n)}$ can be found 
in Refs.~\cite{split12345}.  Substituting 
Eq.~(\ref{eq:msfac}) into Eq.~(\ref{eq:mfxsection}) and 
comparing with Eq.~(\ref{eq:xsection}) we find
\begin{eqnarray}
\label{eq:massfac}
\frac{d \sigma_{ab}^V}{dY}\left(z,u\right) 
=\int_{\sqrt{\frac{z}{u}}}^1 \; dy_1 \; \int_{\sqrt{uz}}^1 \; dy_2 \; 
\Gamma_{ca}(y_1) \;  
\frac{d \tilde{\sigma}_{cd}^V}{dY}
   \left( \frac{z}{y_1 y_2}, \frac{y_1 u}{y_2} \right) \; 
\Gamma_{db}(y_2).
\end{eqnarray}
The convolution integrals follow contours in the $(z,u)$ plane,
as shown in Fig.~\ref{uzplane}.  The $y_1$ integration, holding
$y_2$ fixed, sweeps out a flow such as the one marked ``left collinear'';
whereas the $y_2$ integration sweeps along a ``right collinear'' line.
The lower limits of the integration over the $y_i$ correspond to the point
$(\tilde{z}, \tilde{u}) \equiv ({z \over y_1y_2}, {y_1 \over y_2} u )$
striking one of the two boundaries,
$\tilde{u} = \tilde{z}$ or $\tilde{u} = 1/\tilde{z}$.
We solve Eq.~(\ref{eq:massfac}) for the finite partonic 
cross sections $d\tilde{\sigma}_{ab}/dY$ recursively, order by order 
in the $\alpha_s$  expansion. 

At this point it is straightforward to derive the finite partonic cross
sections.  We outline below the salient features of the calculation.  All
cross sections referred to in the formulas below are finite; we henceforth
drop the superscript tilde when referring to them.  We will also drop
``$d/dY$'' to make the formulas more compact.
\begin{itemize}
{\item To ${\cal O}\left(\alpha_{s}^{2}\right)$, at least one of the two
$\Gamma_{ab}$ factors, or $d \sigma_{ab} /dY$, on the right-hand side of 
Eq.~(\ref{eq:massfac}) has a delta function containing the convolution 
variable.  If neither $\Gamma_{ab}$ factor contains a delta function,
then only the LO cross section enters, with $\tilde{u} = \tilde{z} = 1$.
This forces both $y_1$ and $y_2$ to be set to their lower endpoints, 
$\sqrt{z/u}$ and $\sqrt{zu}$ respectively, so no integral needs to be done.
Apart from this case, the double integral in Eq.~(\ref{eq:massfac}) 
reduces to a single integral of one of the following two forms: 
a ``right'' convolution,
\begin{equation}
\left[ \sigma_{ab} \otimes P^{(n)}_{bc} \right]\left(z,u\right) 
= \int_{\sqrt{uz}}^{1} dx \, \sigma_{ab} 
\left(\frac{z}{x},\frac{u}{x}\right)P^{(n)}_{bc}\left(x\right),
\label{rightconvol}
\end{equation}
or a ``left'' convolution,
\begin{equation}
\left[ P^{(n)}_{ba} \otimes \sigma_{bc} \right]\left(z,u\right) 
= \int_{\sqrt{\frac{z}{u}}}^{1} dx \, \sigma_{ab} 
\left(\frac{z}{x},xu\right)P^{(n)}_{bc}\left(x\right).
\label{leftconvol}
\end{equation}
Using the behavior of the partonic cross sections under inversion 
of rapidity, 
$\sigma_{ab}\left(z,\frac{1}{u}\right)=\sigma_{ba}\left(z,u\right)$, 
it is simple to show that 
\begin{equation}
\left[ \sigma_{ab} \otimes P^{(n)}_{bc} \right]\left(z,\frac{1}{u}\right)
	=\left[ P^{(n)}_{bc} \otimes \sigma_{ba} \right]\left(z,u\right).
\label{convolrel}
\end{equation}
We need only consider right convolutions; we can obtain left convolutions 
by inverting the variable $u$.}
{\item The convolutions required to obtain a 
finite NLO cross section are of the form $\sigma_{ab}^{(0)} \otimes 
P^{(0)}_{bc}$.  The LO cross section has the following form:
\begin{equation}
\sigma_{q\bar{q}}^{(0)} 
\propto \delta\left(1-z\right)\left\{\delta\left(y\right)
+\delta\left(1-y\right)\right\},
\label{borncs}
\end{equation}
where we have used the variable $y$ defined in 
Eq.~(\ref{eq:ydef}).  Substituting $\sigma^{(0)}$ into the convolution 
formula in Eq.~(\ref{rightconvol}), we find that the 
resulting $\delta\left(1-\frac{z}{x}\right)$ removes the 
integration, leaving only the product of 
$P^{(0)}_{bc}\left(z\right)$ with the remainder of 
$\sigma^{(0)}$.  We note that this remainder contains 
either $\delta\left(y\right)$ or $\delta\left(1-y\right)$.
To put it another way, in Eq.~(\ref{rightconvol}), 
since $\sigma_{q\bar{q}}^{(0)}(z/x,u/x)$ requires $z/x=u/x=1$,
the terms generated all have $u=z$, corresponding to $y=0$.
The $\delta\left(1-y\right)$ term only contributes in the limit
of Born kinematics, $u=z=1$.}
{\item There are three distinct types of convolutions needed 
in the NNLO cross section: $\sigma_{ab}^{(0)} \otimes 
P^{(1)}_{bc}$, $\left[ \sigma_{ab}^{(0)} \otimes P^{(0)}_{bc} \right] 
\otimes P^{(0)}_{cd}$, and 
$\sigma_{ab}^{(1)} \otimes P^{(0)}_{bc}$.  
The first of these is simple; as in the NLO cross section, the 
$\delta\left(1-\frac{z}{x}\right)$ from the Born cross section 
removes the convolution integral, and all the terms generated have $u=z$. 
We discuss the remaining cases below in some detail.
\begin{itemize}
\item{We solve the second type iteratively.  
The $\sigma_{ab}^{(0)} \otimes P^{(0)}_{bc}$ piece was already 
computed to obtain the NLO cross section.  It contains either 
$\delta\left(y\right)$ or $\delta\left(1-y\right)$, 
as noted above.  It may also contain distributions in $1-z$.  
It is simple to show that when performing the second convolution 
integral using Eq.~(\ref{rightconvol}), 
$\delta\left(1-y\right) \rightarrow \delta\left(x-\sqrt{uz}\right)$
(and again $u=z$), removing the integration.  
In the $\delta\left(y\right)$ terms, it is convenient to treat plus 
distributions as follows: for distributions of $1-z$, we set 
\begin{equation}
\left[ \frac{{\rm ln}^{n}\left(1-z\right)}{1-z}\right]_{+} 
\rightarrow \left(1-z\right)^{-1+\epsilon} \Bigg|_{\epsilon^n},
\label{zdists}
\end{equation}
where the vertical bar indicates that we should take the appropriate 
term in the $\epsilon$ expansion defined in Eq.~(\ref{eq:plusexpansion}).  
For distributions of $1-x$ arising from the splitting function, we use 
\begin{equation}
\left[\frac{1}{1-x}\right]_{+} 
\rightarrow \left(1-x\right)^{-1+a\epsilon} \Bigg|_{a^0},
\label{splitdist}
\end{equation}
where we now must take the ${\cal O}\left(a^0\right)$ term.  The most 
complicated integral we must evaluate, which contains plus distributions 
in both $1-z$ and $1-x$, becomes
\begin{equation}
I_1=\delta\left(y\right)\int_{z}^{1} dx \, f(z/x) 
\, \left(1-\frac{z}{x}\right)^{-1+\epsilon} 
\left(1-x\right)^{-1+a\epsilon},
\label{examp1a}
\end{equation}
where $f(z/x)$ is finite in all kinematic limits and we have used 
the delta function to simplify the lower limit of integration.  
Performing the variable change $q=(x-z)/(1-z)$, we obtain
\begin{equation}
I_1=\delta\left(y\right)\int_{0}^{1} dq \, 
\left[q(1-z)+z\right]^{1-\epsilon} f(z/x[q]) 
 \left(1-z\right)^{-1+\epsilon\left(1+a\right)}q^{-1+\epsilon}
\left(1-q\right)^{-1+a\epsilon},
\label{examp1b}
\end{equation}
where $x[q] = q(1-z) + z$.
We can extract the distributions in $1-z$ by using the expansion in 
Eq.~(\ref{eq:plusexpansion}).  We must 
also interpret the $q$ and $1-q$ factors as distributions; we set 
\begin{equation}
\label{qexpansion}
q^{-1+\epsilon} = \frac{1}{\epsilon} \delta(q) + \sum_n 
\frac{\epsilon^n}{n!} \left[\frac{\ln^n q}{q} \right]_+,
\end{equation}
and utilize a similar expansion for $1-q$.  We can now expand the 
integrand to ${\cal O}\left(a^0\epsilon^n\right)$.  
Performing the required integrations, we obtain the result for 
this convolution in terms of polylogarithms of rank 2 and 3 
in the variable $z$.}
{\item To obtain convolutions of the form 
$\sigma_{ab}^{(1)} \otimes P^{(0)}_{bc}$, we first return to the 
form of the NLO cross section before expansion in $\epsilon$, which is  
\begin{equation}
\sigma_{q\bar{q}}^{(1)} 
\propto y^{-1-\epsilon}\left(1-y\right)^{-1-\epsilon}
\left(1-z\right)^{-2-2\epsilon} 
	+\ldots \,\, .
\label{NLOcs}
\end{equation}
The ellipsis denotes terms of the form 
$\sigma_{ab}^{(0)} \otimes P^{(0)}_{bc}$, which are needed for an infrared 
finite NLO cross section; the convolution of these with $P^{(0)}_{cd}$ 
has already been discussed, and we ignore them here.  We have presented 
the $q\bar{q}$ cross section; the $qg$ NLO result differs only in the 
exponents of $y$, $1-y$, and $1-z$ which appear, and the required 
convolutions proceed similarly to those we now discuss.  We again 
consider the case where the splitting function contains a plus
distribution in $1-x$.  We rewrite this term using  Eq.~(\ref{splitdist}).  
The integral we must evaluate becomes
\begin{equation}
I_2=\int_{\sqrt{uz}}^{1} dx \, f\left(\frac{z}{x},y_p\right)
\, y_{p}^{-1-\epsilon} \left(1-y_p\right)^{-1-\epsilon} 
\left(1-\frac{z}{x}\right)^{-2-2\epsilon} \left(1-x\right)^{-1+a\epsilon},
\label{examp2a}
\end{equation}
where
\begin{equation}
y_p = \frac{x\left(u-z\right)}{\left(x-z\right)\left(x+u\right)}
\label{ypdef}
\end{equation}
and $f(\frac{z}{x},y_p)$ is finite in all kinematic limits.  
Performing the variable change 
$q=\left(x-\sqrt{uz}\right) /  \left(1-\sqrt{uz}\right)$, 
the integral becomes
\begin{eqnarray}
I_2&=&\int_{0}^{1} dq \, f(z,y;x[q])\,\, g(z,y;x[q],\epsilon) 
\,\, y^{-1-\epsilon} \left(1-y\right)^{-1-\epsilon\left(1-a\right)} 
\left(1-z\right)^{-1-\epsilon} \nonumber \\ 
& & \times q^{-1-\epsilon}\left(1-q\right)^{-1+a\epsilon},
\label{examp2b}
\end{eqnarray}
where $x[q] = q(1-\sqrt{uz}) + \sqrt{uz}$.
We have absorbed terms which are finite in all limits into the function $g$.  
We extract the singularities in $y$, $1-y$, and $1-z$ using the expansion of 
Eq.~(\ref{eq:plusexpansion}); we again interpret the $q$ and $1-q$ factors 
as distributions, and expand them as in Eq.~(\ref{qexpansion}). 
We can now expand the integrand in both $a$ and $\epsilon$.  To obtain 
the contribution to the NNLO cross section, we take the 
${\cal O}\left(a^0 \right)$ term, and expand it in $\epsilon$ up to and 
including the ${\cal O}\left(\epsilon^0 \right)$ piece.  The resulting 
integrals are straightforward to evaluate, and again give polylogarithms 
of ranks 2 and 3.  The rank 3 polylogarithms only appear in the
$\delta(y)$ terms, and are functions of $z$ only.
}
\end{itemize}
}
\end{itemize}

After performing both the UV renormalization and the collinear 
subtractions discussed above, we obtain finite partonic cross sections.

\section{Partonic cross sections}

The basic quantities we compute, $d^2\sigma^{V\to {\rm leptons}}/dM/dY$,
include the probability for the vector boson $V$ to decay into a 
pair of leptons, {\it e.g.} $Z \to l^+l^-$ or $W^+ \to l^+\nu_l$, 
and are differential in both rapidity $Y$ and di-lepton invariant mass $M$.
We shall present our results in a format which is normalized properly for 
virtual photon production, $\gamma^* \to l^+l^-$ (see Eq.~(\ref{channelsum})
below).  For $W$ and $Z$ production, as well as for $\gamma$-$Z$
interference in the $l^+l^-$ channel, we introduce additional normalization 
factors $N^V$, where:
\begin{eqnarray}
N^{\gamma} &=& 1; \nonumber \\ 
N^Z &=& \frac{3}{16 s_{W}^{2}c_{W}^{2}\alpha_{\rm QED}} 
\frac{\Gamma_Z B^{Z}_{l}}{M_Z} \frac{M^4}{(M^2-M_{Z}^2)^2
	+\Gamma_{Z}^{2}M_{Z}^{2}}; \nonumber \\ 
N^{\gamma Z} &=& \frac{v^{\gamma}_{l} v^{Z}_{l}}{8 s_{W}^{2}c_{W}^{2}} 
\frac{M^2(M^2-M_{Z}^{2})}{(M^2-M_{Z}^2)^2
	+\Gamma_{Z}^{2}M_{Z}^{2}}; \nonumber \\ 
N^{W} &=& \frac{3}{4s_{W}^{2}\alpha_{\rm QED}} 
\frac{\Gamma_W B^{W}_{l}}{M_W} \frac{M^4}{(M^2-M_{W}^2)^2
	+\Gamma_{W}^{2}M_{W}^{2}}.
\label{normalizations}
\end{eqnarray}
We have used the notations $\Gamma_{Z}$ and $\Gamma_{W}$ for the total
widths of the $Z$ and $W$, $M_Z$ and $M_W$ for their masses, and
$B^{Z}_{l}$ and $B^{W}_{l}$ for their branching fractions into leptons.
The leptonic vector couplings appearing in $N^{\gamma Z}$ are given by
\begin{equation}
v_l^\gamma = -1, \qquad v_l^Z = -1 + 4 s_{W}^2,
\label{vleptonic}
\end{equation}
and $s_{W}$ and $c_{W}$ represent the sine and cosine of the weak 
mixing angle, respectively.

Finally, we require the luminosity functions
$L^{V}_{ij}\left(x_1,x_2\right)$ that enter the hadronic rapidity
distribution.  These functions contain the PDFs for the partons $i,j$, 
and appropriate combinations of the electroweak couplings to $V$.  
We follow closely the notation of Ref.~\cite{neerven}.  
We first introduce the following $2n_f \times 2n_f$ matrices:
\begin{eqnarray}
C^{ii}_{\gamma ,Z}\left(q_k,q_l\right) 
= C^{ff}_{\gamma ,Z}\left(q_k,q_l\right) &=& 
	\left\{ \begin{array}{ll} 1 & \mbox{if $q_k=\bar{q}_l$} \\
				  0 & \mbox{otherwise}
	        \end{array} \right. ; \nonumber \\ 
C^{if}_{\gamma ,Z}\left(q_k,q_l\right) &=& 
	\left\{ \begin{array}{ll} 1 & \mbox{if $q_k=q_l$} \\
				  0 & \mbox{otherwise}
	        \end{array} \right. ; \nonumber \\
C^{ii}_{W^{\pm}}\left(q_k,q_l\right) &=&
	\left\{ \begin{array}{ll} |V_{q_kq_l}|^2 
     & \mbox{if $e_{q_k}+e_{q_l}=\pm 1$} \\
				  0 & \mbox{otherwise}
	        \end{array} \right. ; \nonumber \\ 
C^{if}_{W^{\pm}}\left(q_k,q_l\right) &=&
	\left\{ \begin{array}{ll} |V_{q_kq_l}|^2 
     & \mbox{if $e_{q_k}=\pm 1 +e_{q_l}$} \\
				  0 & \mbox{otherwise}
	        \end{array} \right. ; \nonumber \\ 
C^{ff}_{W^{\pm}}\left(q_k,q_l\right) &=&
	\left\{ \begin{array}{ll} |V_{q_kq_l}|^2 
     & \mbox{if $e_{q_k}+e_{q_l}=\mp 1$} \\
				  0 & \mbox{otherwise}
	        \end{array} \right. .
\label{Cmatrices}
\end{eqnarray}
Here $q_k$ is an element of either of the following $n_f$-dimensional vectors: 
$Q=\left\{u,d,s,c,b \right\}$,
$\bar{Q}=\left\{\bar{u},\bar{d},\bar{s},\bar{c},\bar{b}\right\}$.  
In Eq.~(\ref{Cmatrices}), $e_{q_k}$ denotes the electric charge of 
the element, 
and $V_{q_k q_l}$ indicates the appropriate CKM matrix element.  
Using these matrices, we can write the luminosity functions as follows:
\begin{eqnarray}
L^{V}_{NS}\left(x_1,x_2\right) 
   &=& \sum_{i,j \in Q,\bar{Q}} C^{ii}_{V}\left(q_i,\bar{q}_j\right) 
  \left( v^{V,2}_{i}+a^{V,2}_{i}\right)
 \,q_{i}\left(x_1\right)\,\bar{q}_{j}\left(x_2\right); \nonumber \\
L^{V}_{B^2}\left(x_1,x_2\right) 
   &=& \sum_{i \in Q,\bar{Q}}\, \sum_{k,l \in Q} 
         C^{ff}_{V}\left(q_k,\bar{q}_l\right) 
	\left( v^{V,2}_{k}+a^{V,2}_{k}\right)
       \,q_{i}\left(x_1\right)\,\bar{q}_{i}\left(x_2\right); \nonumber \\
L^{V}_{BC}\left(x_1,x_2\right) 
   &=& \sum_{i \in Q,\bar{Q}}\, \sum_{k \in Q,\bar{Q}} 
	\left[C^{if}_{V}\left(q_i,\bar{q}_k\right)
             +C^{if}_{V}\left(\bar{q}_i,q_k\right)\right]
	\left( v^{V,2}_{i}+a^{V,2}_{i}\right)
        \,q_{i}\left(x_1\right)\,\bar{q}_{i}\left(x_2\right); \nonumber \\
L^{V}_{AB,vec}\left(x_1,x_2\right) 
   &=& \sum_{i \in Q,\bar{Q}}\, \sum_{k \in Q} 
              C^{ff}_{V}\left(q_k,\bar{q}_k\right) v^{V}_{i}v^{V}_{k}
        \,q_{i}\left(x_1\right)\,\bar{q}_{i}\left(x_2\right); \nonumber \\ 
L^{V}_{AB,ax}\left(x_1,x_2\right) 
   &=& \sum_{i \in Q,\bar{Q}}\, \sum_{k \in Q} 
        C^{ff}_{V}\left(q_k,\bar{q}_k\right) a^{V}_{i}a^{V}_{k}
        \,q_{i}\left(x_1\right)\,\bar{q}_{i}\left(x_2\right); \nonumber \\ 
L^{V}_{qg}\left(x_1,x_2\right)
   &=& \sum_{i,j \in Q,\bar{Q}} C^{if}_{V}\left(q_i,q_j\right) 
	\left( v^{V,2}_{i}+a^{V,2}_{i}\right)
        \,q_{i}\left(x_1\right)\,g\left(x_2\right); \nonumber \\
L^{V}_{gq}\left(x_1,x_2\right)&=&L^{V}_{qg}\left(x_2,x_1\right); \nonumber \\
L^{V}_{C^2}\left(x_1,x_2\right)
   &=& \sum_{i,j \in Q,\bar{Q}}\, \sum_{k \in Q,\bar{Q}} 
       C^{if}_{V}\left(q_i,q_k\right) \left(v^{V,2}_{i}+a^{V,2}_{i}\right)
        \,q_{i}\left(x_1\right)\,q_{j}\left(x_2\right); \nonumber \\
L^{V}_{D^2}\left(x_1,x_2\right)
   &=& \sum_{i,j \in Q,\bar{Q}}\, \sum_{k \in Q,\bar{Q}} 
       C^{if}_{V}\left(q_j,q_k\right) \left(v^{V,2}_{j}+a^{V,2}_{j}\right)
         \,q_{i}\left(x_1\right)\,q_{j}\left(x_2\right); \nonumber \\
L^{V}_{CD,vec}\left(x_1,x_2\right)
   &=& \sum_{i,j \in Q,\bar{Q}}\, \sum_{k \in Q,\bar{Q}} 
	C^{if}_{V}\left(q_i,q_i\right) v^{V}_{i}v^{V}_{j} 
     \,q_{i}\left(x_1\right)\,q_{j}\left(x_2\right); \nonumber \\
L^{V}_{CD,ax}\left(x_1,x_2\right)
   &=& \sum_{i,j \in Q,\bar{Q}}\, \sum_{k \in Q,\bar{Q}} 
	C^{if}_{V}\left(q_i,q_i\right) a^{V}_{i}a^{V}_{j} 
     \,q_{i}\left(x_1\right)\,q_{j}\left(x_2\right); \nonumber \\
L^{V}_{CE_1}\left(x_1,x_2\right) &=& \sum_{i,j \in Q,\bar{Q}} 
     C^{if}_{V}\left(q_i,q_j\right) \left( v^{V,2}_{i}+a^{V,2}_{i}\right)
      \,q_{i}\left(x_1\right)\,q_{j}\left(x_2\right); \nonumber \\
L^{V}_{CE_2}\left(x_1,x_2\right)
  &=& L^{V}_{CE_1}\left(x_2,x_1\right); \nonumber \\
L^{V}_{CF}\left(x_1,x_2\right)
  &=& \sum_{i \in Q,\bar{Q}}\,  \sum_{j \in Q,\bar{Q}}
	 C^{if}_{V}\left(q_i,q_j\right) \left( v^{V,2}_{i}+a^{V,2}_{i}\right)
	\,q_{i}\left(x_1\right)\,q_{i}\left(x_2\right); \nonumber \\
L^{V}_{gg}\left(x_1,x_2\right)
  &=& \sum_{i,j \in Q} C^{ff}_{V}\left(q_i,\bar{q}_j\right) 
	\left( v^{V,2}_{i}+a^{V,2}_{i}\right)
        \,g\left(x_1 \right)\, g\left(x_2 \right).
\label{luminosities}
\end{eqnarray}
In this formula, a function such as $q_i\left(x_1 \right)$ denotes the 
appropriate parton distribution function.  The label $V$ takes the 
values $\gamma$, $Z$, and $W^{\pm}$.  The electroweak couplings
$v_i^V$ and $a_i^V$ are given in Eq.~(\ref{vadefs}).
To obtain the $\gamma$-$Z$ interference luminosity 
functions, we must use $V=\gamma$ and substitute 
$v^{\gamma ,2}_{i} \rightarrow v^{\gamma}_{i}v^{Z}_{i}$, 
$v^{\gamma}_{i}v^{\gamma}_{j} \rightarrow 
\frac{1}{2}\left( v^{\gamma}_{i}v^{Z}_{j}+v^{\gamma}_{j}v^{Z}_{i}\right)$.

The final ingredients required are the partonic hard cross sections
for the channels corresponding to the luminosity 
functions~(\ref{luminosities}), $d\sigma_{ij}^V/dY(z,u)$ for
$ij \in \{ NS, B^2, BC, \ldots, gg \}$.
We have obtained analytic expressions for these functions; however, they 
are quite lengthy, so we refrain from giving them here.  
A MAPLE file containing the functions is available from the authors by 
request.  They have also been implemented in C++, 
as part of a numerical program computing the hadronic rapidity distribution.
The bulk of the analytical complexity stems from the ``hard'' region,
away from the boundaries, $z < 1$ and 
$z < u < 1/z$ (or $0 < y < 1$).  

The hard functions contain polylogarithms of rank 2, $\Li_2(A_i(z,u))$, 
and there are a large number of possible ways the arguments $A_i$ can 
depend on the underlying variables $z,u$.
In most cases, the arguments are rational functions of $t = \sqrt{z}$
and $r = \sqrt{u}$, as in the case of the sample integral ${\cal J}(z,u)$
presented in Eqs.~(\ref{eq:A0zu}) and (\ref{eq:f0u}).
In four cases, though, we have to introduce functions in which the
polylogarithmic arguments are significantly more complicated.  
The four functions of this type, $J_3$, $J_{27}$, $J_{21}$ and $J_2$,
are given by:
\begin{eqnarray}
J_3(z,u) &=& { 1 \over 1+u } \Re \biggl[ - {1\over 4} \ln^2(z/u)
     - {1\over 4} \ln(1+u) \biggl( 
                \ln(z/u) + 2 \ln(1+t r) 
\nonumber \\
&& \hskip3cm
             - 2 \ln\biggl( { d_1 - r - 2 t (1+u) \over d_1 - r } \biggr)
             - 2 \ln\biggl( { d_1 + r + 2 t (1+u) \over d_1 + r } \biggr)
              \biggr)
\nonumber \\
&&  \hskip1cm
     + \Li_2\biggl( { 2 t (1+u) \over d_1 - r } \biggr) 
     + \Li_2\biggl( { - 2  t (1+u) \over d_1 + r } \biggr)
\nonumber \\
&&  \hskip1cm
     - \Li_2\biggl( { 2 z (1 +u) \over r(d_1 - r) } \biggr)
     - \Li_2\biggl( { - 2 z (1 +u) \over r(d_1 + r) } \biggr) \biggr] \,,
\label{J3}
\end{eqnarray}
where $d_1 = \sqrt{u+4z(1+u)}$;
\begin{eqnarray}
J_{27}(z,u) &=& - {1 \over 2 r d_1 } \Re \biggl[
	 \ln\biggl( { r_2 - r_1 + r_1 r_2 - 3 - 2 u
                \over r_2 - r_1 - r_1 r_2 + 3 + 2 u } \biggr)
\nonumber \\
&&  \hskip1cm
     +  \ln\biggl( { d_1 - 2 t r_1 + r_1 r - r_1 d_1 + 2 t + 2 t u + r
	       \over d_1 - 2 t r_1 + r_1 r + r_1 d_1 - 2 t - 2 t u - r } 
            \biggr)
\nonumber \\
&&  \hskip1cm
         + 3 \ln\biggl( { (r - d_1 + 2 t r_1) (1 + r_2 - 2 r_1)
                    \over (r + d_1 - 2 t r_1) (1 - r_2 + 2 r_1) } \biggr)
   \biggr] \,,
\label{J27}
\end{eqnarray}
where $r_1 = \sqrt{1+u}$, $r_2 = \sqrt{5+4u}$;
\begin{eqnarray}
J_{21}(z,u) &=& \Re \Biggl\{  { i \over z (1+u) x_1 } \biggl[
      - \ln(t a_1^+) \ln\biggl( { a_1^+ - 1 \over 1 - t r a_1^+ } \biggr)
      + \ln(t a_1^-) \ln\biggl( { a_1^- - 1 \over 1 - t r a_1^- } \biggr)
\nonumber \\
&&  \hskip1cm
      - \Li_2(1 - a_1^+) + \Li_2(1 - t r a_1^+) 
      + \Li_2(1 - a_1^-) - \Li_2(1 - t r a_1^-) \biggr] \Biggr\}
\nonumber \\
&&  \hskip0.2cm
 + { (1+z) (1 - t r ) \over (1-z)^2 (1+u) r (r+t) } 
       \ln z\ \ln\biggl( { 1+z \over 2 } \biggl) \,,
\label{J21}
\end{eqnarray}
where 
\begin{eqnarray}
 x_1 &=& \sqrt{ \biggl({ 2 u \over 1+u } \biggr)^2 {1\over z} - 1 },
    \qquad z \leq \biggl({ 2 u \over 1+u } \biggr)^2 \,,
\nonumber \\
  &=& i \sqrt{ 1 - \biggl({ 2 u \over 1+u } \biggr)^2 {1\over z} }, 
    \qquad z > \biggl( { 2 u \over 1+u } \biggr)^2 \,,
\label{x1def}
\end{eqnarray}
\begin{equation}
a_1^\pm = { 2 u  \over z (1+u) (1 \pm ix_1) } \,;
\label{a1pmdef}
\end{equation}
and 
\begin{eqnarray}
J_2(z,u) &=& - J_3(z,u) 
  +  { 1 \over 1+u } \biggl\{
    - r d_1 \, \ln\biggl| { 2 + u + r d_1 \over 2 + u - r d_1 } \biggr| 
            \, J_{27}(z,u)
\nonumber \\
&&  \hskip0.2cm
 + {3\over 4} \Re \biggl[  
    l_1
    \ln\biggl| { u (u+2) (r_2 - 1) \over (1+u) (2 + u - u r_2) } \biggr|
   - \ln(1+u) \ln\biggl| { (2 + u - u r_2) (r + d_1)
                         \over  r (1+r_2) (2 + u - r d_1) } \biggr|
\nonumber \\
&&  \hskip1.2cm
   - l_2
     \ln\biggl( { r (u+2 ) (d_1 - r) \over (1+u) (2 + u - r d_1) } \biggr)
   - {1\over6} ( l_1^2 - l_2^2 )
\nonumber \\
&&  \hskip1.2cm
   + \Li_2\biggl( { - u (u+2) (1 + r_2) \over 2 + u - u r_2 } \biggr) 
   - \Li_2\biggl( { - r (u+2) (r + d_1) \over 2 + u - r d_1 } \biggr)
\nonumber \\
&&  \hskip1.2cm
   - \Li_2\biggl( { 2 + u + r d_1 \over (1+u) (2 + u - r d_1) } \biggr)
   + \Li_2\biggl( { 2 + u + u r_2 \over (1+u) (2 + u - u r_2) } \biggr)
    \biggr]
\nonumber \\
&&  \hskip0.2cm
 + {1\over 2} \Bigl( M_1(x_u) - M_1(x_l) \Bigr) 
 + {3\over8} \pi^2 \, \Theta(u-1) \biggr\} \,,
\label{J2}
\end{eqnarray}
where $\Theta(x)$ is the Heaviside function,
\begin{eqnarray}
l_1 &=& \ln\biggl| { 2 + u + u r_2 \over 2 + u - u r_2 } \biggr| \,,
\qquad 
l_2 = \ln\biggl| { 2 + u + r d_1 \over 2 + u - r d_1 } \biggr| \,,
\nonumber \\
M_1(x) &=& \Re \Biggl[ 
   - \Li_2\biggl( { x+x^+ \over 2 x^+ } \biggr)
   + \Li_2\biggl( { x+x^- \over 2 x^- } \biggr)
   - \Li_2\biggl( { x^+ - x \over x^+ - x^- } \biggr)
   + \Li_2\biggl( { x + x^+ \over x^+ - x^- } \biggr)
\nonumber \\ 
&& \hskip0.2cm
   + {1\over4} \biggl( \ln^2\biggl( { x + x^+ \over x - x^+ } \biggr)
                     - \ln^2\biggl( { x + x^- \over x - x^- } \biggr) \biggr)
   - {1\over2}  \biggl( \ln^2\biggl( { x+x^+ \over 2 x^+ } \biggr)
                      - \ln^2\biggl( { x+x^- \over 2 x^- } \biggr) \biggr)
\nonumber \\ 
&& \hskip0.2cm
   + {1\over2} \ln\biggl( { x + x^+ \over x - x^+ } \biggr)
          \ln\biggl( {(x+x^-)(x-x^-) \over (x^+-x^-)^2} \biggr)
             \Biggr] \,,
\nonumber \\
x^+ &=&  - { 2 + u - 2 r_1 \over u } = { 1\over x^- } \,,
\qquad
x_u = r_2 + 2 r_1, 
\qquad
x_l = { d_1 + 2 \sqrt{z (1 +u)} \over r } \,.
\label{lMxdefs}
\end{eqnarray}

After the use of polylogarithmic identities, the set of arguments
of the remaining polylogarithms, $\Li_2(A_i)$, can be reduced,
if desired, to
\begin{eqnarray}
A_i &\in& \biggl\{ -z, z, { 1+z \over 2 }\,, -r, -u, { 1-u \over 1+u } \,,
 { 1-u \over 2 }\,, -{t\over r} \,, {t\over r} \,, { r - t \over 2r }\,,
  -tr, tr, { 1 - tr \over 2}\,,  { 1 - r \over 1 + tr }\,,
 { r - 1 \over r + t } \,, 
\nonumber \\ 
&& \hskip0.3cm
 { 1 - t \over r + 1 } \,, { r (1-t) \over r + 1 }\,,  
 { u - z \over 1 + u } \,, { 1 - uz \over 1 + u } \,, 
 { 1 - tr \over 1 + u } \,, { r (r-t) \over 1 + u } \,,
 { u - 1 \over r (r+t) } \,, { 1- u \over 1 + tr } \biggr\} \,.
\label{Aiset}
\end{eqnarray}
The arguments of the logarithms that appear, $\ln(B_i)$, are drawn 
from a simpler set,
\begin{eqnarray}
B_i &\in& \{ z, 1-z, 1+t, 1+z, u, r-1, r+1, 1+u, 
                   r-t, r+t, 1-tr, 1+tr \},
\label{Biset}
\end{eqnarray}
but since they can appear in pairs, there are still quite a few
terms of the form $\ln(B_i)\ln(B_j)$.

As mentioned above, rank 3 polylogarithms of a single variable,
$z$, are generated in the collinear regions, $u=z$ ($\delta(y)$ terms)
and $u=1/z$ ($\delta(1-y)$ terms).  These collinear
terms have a similar form to the NNLO total cross section, integrated 
over rapidity~\cite{neerven,HarlanderKilgore}.
We can write the functions appearing, $\Li_3(a_i)$, in terms of
\begin{eqnarray}
a_i &\in& \biggl\{ z, -z, 1-z,  -1-z, 1-z^2, { 1 + z \over 2 }\,, 
{ 1 - z \over 2 }\,, { 1 - z \over 1 + z }\,, { 2 z \over 1 + z } \,,
\nonumber \\ 
&& \hskip0.3cm
-{ z \over2 } \,,
{ z \over 2 (1+z) } \,, -{ 1 \over 2 z } \,, { 1 \over 2(1+z) } \,,
- { 1 + z \over z } \biggr\}.
\label{aiset}
\end{eqnarray}
The rank 2 polylogarithms appearing in the collinear terms, $\Li_2(b_i)$, 
have arguments
\begin{equation}
b_i \in \biggl\{ z, -z, -1-z, { 1 + z \over 2 }\,, -{ z \over2 } \,,
-{ 1 \over 2 z } \,, - { 1 + z \over z } \biggr\},
\label{biset}
\end{equation}
while the arguments of the logarithms, $\ln(c_i)$, are
\begin{equation}
c_i \in \{ z, 1-z, 1+z, 2+z, 1+2z \}.
\label{ciset}
\end{equation}

\begin{figure}[htbp]
\vspace{0.0cm}
\centerline{
\psfig{figure=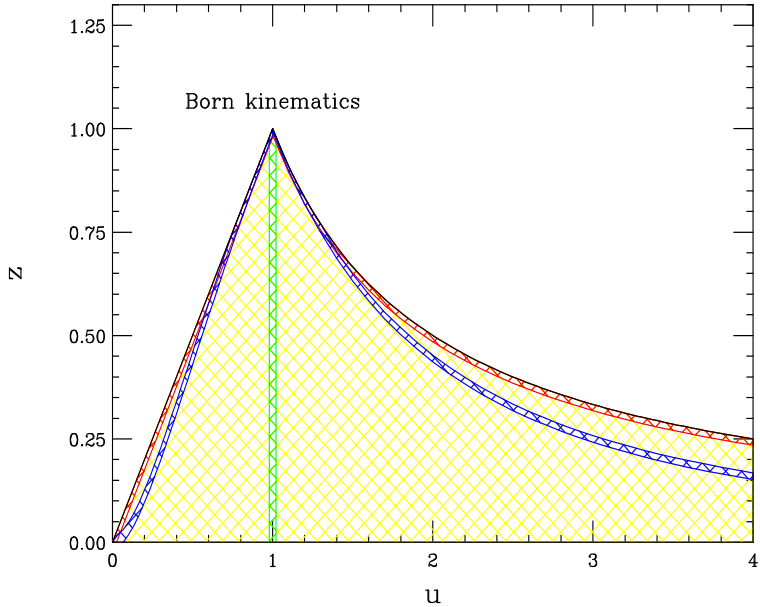,height=8.0cm,width=10.0cm,angle=0}}
\caption{Regions in the $(z,u)$ plane for which the hard functions
have to be patched, because of singular behavior.
Besides the soft limit $z \to 1$, and the left and right collinear edges,
there are spurious singularities as $u\to1$, 
and as $z \to [2u/(1+u)]^2$ and $z \to [2/(1+u)]^2$.}
\label{uzpatch}
\end{figure}

The hard functions have integrable, logarithmic singularities in the soft
limit $z \to 1$ and the collinear limits $u \to z$ and $u\to 1/z$.
However, the complexity of the analytical formulas is such that many of
the individual terms in the hard functions have much more severe
singularities in these limits, {\it e.g.} several powers of $1/(1-z)$ as
$z \to 1$.  These spurious singularities lead to unacceptable roundoff
error. For this reason we construct patching functions, which are used
instead of the full functions in thin strips near the singular regions.
The patching functions are typically constructed by taking the appropriate
limits analytically.  Fig.~\ref{uzpatch} shows the regions in the $(z,u)$
plane which have to be patched.  In addition to the soft and collinear
regions, there are two other types of regions where the singularities are
completely unphysical.  For $z = [2u/(1+u)]^2$, the variable $x_1$ in
Eq.~(\ref{x1def}) vanishes, leading to a singularity in functions
containing $J_{21}(z,u)$.  There is an equivalent singularity at $z =
[2/(1+u)]^2$ in functions containing $J_{21}(z,1/u)$.  In this pair of
strips, the true function is smooth enough that an analytic patch is not
necessary; instead, when the point $(z,u)$ lies in the strip, we replace
its value by the average of two nearby values on either edge of the
strip~\cite{DKOSZ}.  Finally, the limit $u\to1$ is singular, as indicated
by the presence of $(r-1)$ in the set $B_i$ in Eq.~(\ref{Biset}); there
are spurious power-law singularities as well in this limit.

\noindent

The expansion of the hard functions in a series about $z=1$ can be carried
out to very high order, and produces an approximation to the integrand
which is free of spurious singularities.  Working to order $(1-z)^{25}$
results in an expression whose accuracy is completely adequate for
predictions for typical fixed-target kinematics~\cite{gammastar} and for
$W$ and $Z$ production at the Tevatron.  However, for the case of $W$ and
$Z$ production at the LHC, the small value of 
$\tau = M_V^2/S \approx 4 \times 10^{-5}$ means that values of 
$z \sim 0.01$ are actually relevant in the numerical integration.  
We have therefore used the exact, unexpanded, representations of the 
hard functions (plus patches) in order to get sufficient accuracy
for the case of the LHC.

\label{sec:results}

\section{Numerical results}
\label{sec:numerics}
In this section we present numerical results for the $W$ and $Z$ rapidity 
distributions at both the Tevatron and the LHC.  We use the following 
parameters:
$M_Z = 91.1876~{\rm GeV}$,
$\Gamma_Z = 2.4952~{\rm GeV}$, 
$B^{Z}_{l} = 0.03363$,
$M_W = 80.426~{\rm GeV}$, 
$\Gamma_W = 2.118~{\rm GeV}$, and 
$B^{W}_{l} = 0.1082$.
We use the $Z$-pole value of
$\alpha_{\rm QED}(M_z) = 1/128$ for the fine structure constant, and set
${\rm sin}^2{\theta_W} = 0.23143$, the effective mixing angle measured in
$Z$ pole asymmetries at LEP and SLC~\cite{Zpoleasym}.  We expect that
these choices account for the bulk of the factorizable electroweak
radiative corrections, which dominate for nearly resonant production of
$W$ and $Z$ bosons.  A more accurate description would require a
consistent accounting of the electroweak corrections~\cite{WZewkcorrs}.

We also need the following values of the CKM matrix elements to compute 
the $W$ cross section:
\begin{equation}
|V_{ud}| = 0.975,~~~|V_{us}|=0.222,~~~|V_{cd}|=0.222,~~~|V_{cs}|=0.974.
\label{CKMvalues}
\end{equation}
The absolute values of the other matrix elements are obtained by requiring
unitarity of the CKM matrix.  Because the collider center-of-mass energy
is large, it is possible in principle to produce top quarks in
association with the $W$ or $Z$; however, since these processes can be
distinguished experimentally, we exclude them from consideration.  
We also omit top quarks from the virtual corrections, and set the number
of light (massless) quark flavors $n_f$ to 5 in all numerical results 
in this paper. At one loop, the partonic subprocesses 
$q\bar{q} \rightarrow Zg$ and $qg \rightarrow Zq$ include triangle graphs, 
weighted by the axial couplings $a_q^Z$ for the quarks circulating in 
the loop.  For massless quarks, these contributions cancel generation
by generation.  The effect of a finite top quark mass on the $t-b$ 
contribution has been studied previously, and found to be negligibly
small~\cite{neerven,Dicus}, so we omit it here.  

In the previous sections we discussed how the rapidity distributions of
electroweak bosons in partonic collisions can be computed. To obtain
results for hadronic collisions, we must convolute the partonic
differential cross sections with parton distribution functions which
describe the probability of finding a parton with a given 
momentum fraction inside the hadron. The corresponding formula reads:
\begin{equation}
\frac {d^2\sigma^V}{dM dY} = 
\frac{4\pi\alpha^{2}_{\rm QED}}{9 M^3} \sum_{ij} 
\int dx_1 dx_2 \, N^{V} L^{V}_{ij}\left(x_1,x_2\right)
\frac{d\sigma_{ij}^V}{dY} (x_1,x_2),
\label{channelsum}
\end{equation}
where $d\sigma_{ij}^V/dY$ is the partonic cross section, 
$N^{V}$ is the normalization factor for the 
boson $V$, and $L^{V}_{ij}\left(x_1,x_2\right)$ is the corresponding 
luminosity function; these were discussed in the previous section.  
There are three observable cross sections: 
production of a $W^+$; production of a $W^-$; and neutral current 
production of a lepton pair $l^+l^-$, which receives contributions 
from both $\gamma$ and $Z$ exchange as well as from $\gamma$-$Z$ 
interference.

It is convenient to change the integration variables in the above 
formula and express the integration over $x_1$ and $x_2$ through the 
partonic variables $z$ and $y$.  Consider the case of negative rapidity $Y$; 
the results for $Y > 0$ can be obtained by substituting 
$Y \to -Y$ in the formulae below. 
For $Y<0$, using the relations~(\ref{eq:u_def}), (\ref{eq:z_def}),
(\ref{eq:x1x2invert}) and (\ref{eq:ydef}), the integration over 
$x_1$ and $x_2$ in Eq.~(\ref{channelsum}) can be rewritten as:
\begin{eqnarray}
&&\int dx_1 dx_2 \, N^{V} L^{V}_{ij}\left(x_1,x_2 \right)
 \frac{d\sigma_{ij}^V}{dY}(x_1,x_2) = 
\int \limits_{\sqrt{\tau}e^{-Y}}^{1}dz 
\int \limits_{0}^{1} dy\ F_{ij}(z,y)
\nonumber \\
&& \hskip3.5cm
+
\int \limits_{\sqrt{\tau}e^Y}^{\sqrt{\tau}e^{-Y}}dz 
\int \limits_{y_1(z)}^{1} dy\ F_{ij}(z,y)
+
\int \limits_{\tau}^{\sqrt{\tau}e^Y}dz 
\int \limits_{y_1(z)}^{y_2(z)} dy\ F_{ij}(z,y),
\label{zyintegral}
\end{eqnarray}
where
\begin{eqnarray}
&& F_{ij}(z,y) = J(z,y) 
(1-z) \frac{d\sigma_{ij}^V(z,y)}{dY}
N^{V} L^{V}_{ij}\left(x_1,x_2 \right),
\nonumber \\ 
&&
J(z,y) = \frac{\tau (1+z)}{2 z^2 (1-y(1-z))(z+y(1-z))},
\nonumber \\
&& x_1= e^Y \sqrt{\frac{\tau}{z}\frac{(z+y(1-z))}{(1-y(1-z))}},
\qquad
x_2 = e^{-Y} \sqrt{\frac{\tau}{z} \frac{(1-y (1-z) )}{(z+y(1-z))}},
\nonumber \\
&&
y_1(z) = \frac{\tau e^{-2Y}-z^2}{(z+\tau e^{-2Y})(1-z)},
\qquad
y_2(z) = \frac{z(e^{-2Y} - \tau)}{(\tau+e^{-2Y}z)(1-z)}.
\end{eqnarray}
This representation is convenient for numerical integration. 

We now present results for the $W$ and $Z$ rapidity distributions.  For
the NNLO calculations, we use the corresponding set of MRST parton
distribution functions.  The MRST code contains four different sets of
PDFs.  As mentioned in the inroduction, the complete NNLO evolution
kernels needed for a consistent extraction of PDFs at NNLO are not yet
known.  The MRST program contains both the fastest and slowest possible
perturbative evolutions, based upon the known moments of the required
DGLAP equations.  A third set allows an evolution between these extremes.
Finally, a fourth PDF set which seems preferred by large $E_T$ jet
production at the Tevatron is included.  Unless stated otherwise, we use
{\sf mode} 1 of the MRST NNLO PDF code, which corresponds to the
intermediate rate of evolution.

For the most part, we present double-differential cross sections, 
including the decay to leptons,
\begin{equation}
{ d^2\sigma^{V\to {\rm leptons}} \over dM dY } \,.
\label{basicresult}
\end{equation}
For the case of on-shell
vector bosons, these are evaluated at the resonance peak, $M=M_W$ 
or $M=M_Z$.  Of course any experiment will integrate over the resonance
profile.  If this integral is performed in the narrow-resonance
approximation, and if the $\gamma$ exchange and $\gamma$-$Z$ interference
terms are neglected in the case of the $Z$, the result is
\begin{equation}
 { d\sigma^V \over dY } \ B^{V}_{l} 
 = {\pi \over 2} \Gamma_V \times
    { d^2\sigma^{V\to {\rm leptons}} \over dM dY } \biggr|_{M=M_V} \,.
\label{narrowresint}
\end{equation}
The narrow-resonance conversion factor, $\pi \Gamma_V/2$, numerically
evaluates to 3.919~GeV for the $Z$ boson, and 3.327~GeV for the $W$.  
One can further integrate Eq.~(\ref{narrowresint}) over the rapidity $Y$ to
obtain the theoretical prediction for the ``total cross section times
branching ratio,'' $\sigma^V \times B^{V}_{l}$.  Our total cross section
results for the MRST PDFs, for example, agree with results obtained using
the numerical program of Ref.~\cite{neerven}, after we omit $b$ quarks
from the initial state~\cite{MRST2001,StirlingPrivate}.  
(We note that Eqs.~(B.13) and (B.16) in the article in Ref.~\cite{neerven} 
are missing a factor of $T_{\rm f} = {1\over2}$, and the ``103''
at the end of Eq.~(B.11) should have an $x$ multiplying it. 
Also the normalization of the $W$ cross section in Eqs.~(A.3) and (A.11)
should be a factor of 2 larger.  All these factors are properly included 
in the numerical program~\cite{neerven}.)
Our program is also capable of integrating over a range of di-lepton 
invariant masses, without making the narrow-resonance approximation, 
and we shall present one such plot below.

\begin{figure}[htbp]
\vspace{0.0cm}
\centerline{
\psfig{figure=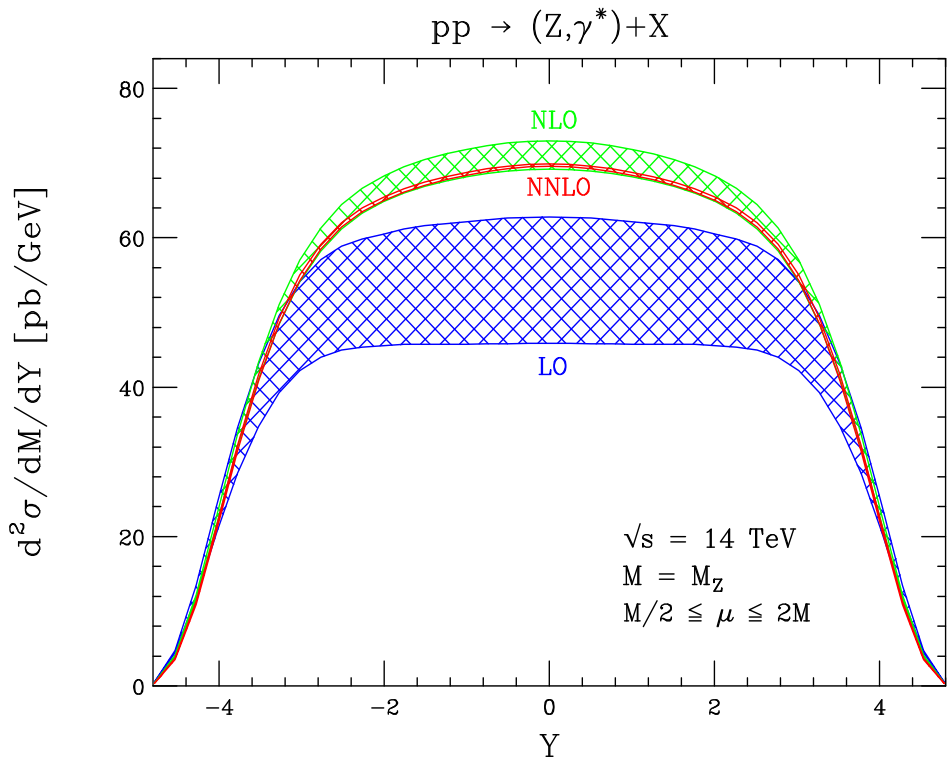,height=10.0cm,width=12.5cm,angle=0}}
\caption{The CMS rapidity distribution of an on-shell $Z$ boson at the
LHC.  The LO, NLO, and NNLO results have been included.  The bands 
indicate the variation of the renormalization and factorization scales 
in the range $M_Z/2 \leq \mu \leq 2M_Z$.}
\label{LHC_Z_Mz}
\end{figure}

We first present, in Fig.~\ref{LHC_Z_Mz}, the rapidity distribution for a
$Z$ boson produced on-shell at the LHC.  The LO, NLO, and NNLO results
have been included.  We have equated the renormalization and factorization
scales, and have varied them in the range $M_Z/2 \leq \mu \leq 2M_Z$.  At
LO the scale variation is large, ranging from 30\% at central rapidities
to 25\% at $Y \approx 3$.  This is reduced to $\approx 6\%$ at NLO for all
rapidities.  At NNLO, the prediction for central rapidities stabilizes
dramatically; the scale variation is $\approx 0.6\%$.  This increases to
1\% at $Y \approx 3$, and 3\% at $Y \approx 4$.  However, it seems that
for $Y \leq 3$, the rapidity values accessible in LHC experiments, the
residual scale dependence is no longer a significant theoretical
uncertainty when the NNLO corrections are included.

The magnitude of the higher-order corrections exhibits a pattern similar
to that of the scale variation.  The NLO corrections significantly
increase the LO prediction; the LO result is increased by 30\% at central
rapidities, and by 15\% for larger rapidity values.  They also change the
shape of the distribution, creating a broad peak at central rapidities, as
is visible in Fig.~\ref{LHC_Z_Mz}.  The results stabilize completely at
NNLO.  The NNLO corrections decrease the NLO result by only 1--2\%, and do
not affect the shape of the distribution.

\noindent

\begin{figure}[htbp]
\vspace{0.0cm}
\centerline{
\psfig{figure=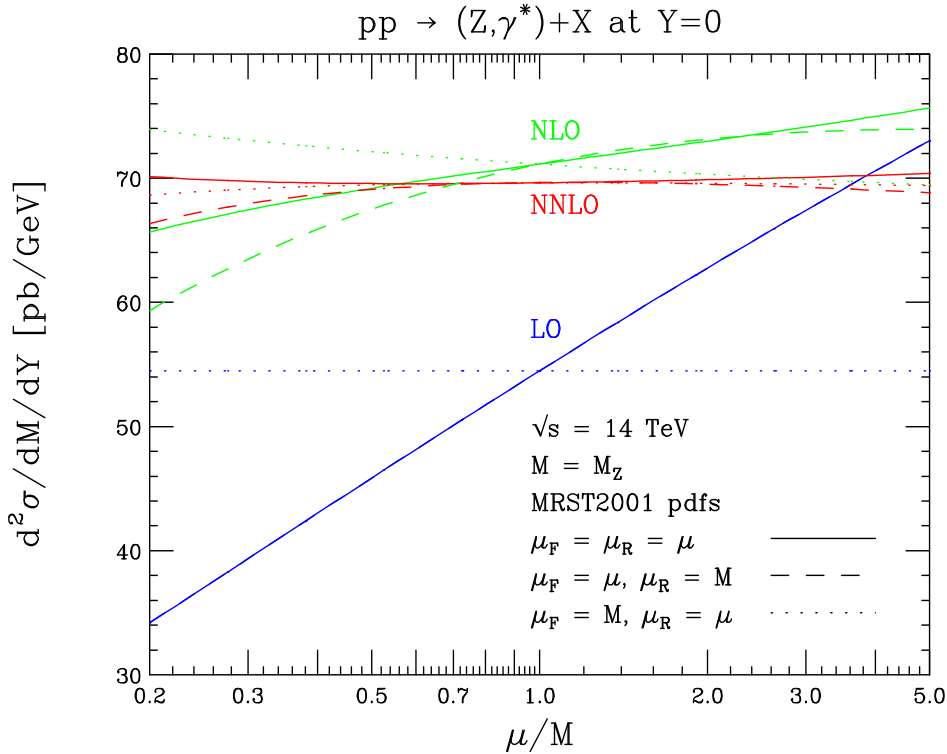,height=10.0cm,width=12.5cm,angle=0}}
\caption{More general variations of the renormalization and factorization
scales, for production of an on-shell $Z$ boson at the LHC, at central
rapidity $Y=0$.   For each order in perturbation theory (LO, NLO, NNLO),
three curves are shown.  The solid curves depict common variation of the 
renormalization and factorization scales, $\mu_F = \mu_R = \mu$,
as used in the rest of the paper, but extending the range of variation
to $M/5 < \mu < 5M$.   The dashed curves represent variation of the 
factorization scale alone, holding the renormalization scale fixed at $M$.
The dotted curves result from varying the renormalization scale instead, 
holding the factorization scale fixed at $M$.}
\label{LHCZy0mu}
\end{figure}

For most of the plots in the paper, in order to estimate the uncertainties 
in the NNLO predictions we shall continue to set $\mu_F = \mu_R = \mu$ and
vary the common scale $\mu$ from $M/2$ to $2M$.  However, it is useful 
to consider a broader range of scale variations, for at least one 
kinematic configuration.  In Fig.~\ref{LHCZy0mu} we study dependence
on $\mu_F$ and $\mu_R$ in more detail for the case of on-shell $Z$ boson
production at the LHC, at the precisely central rapidity point $Y=0$.  
For each order in perturbation theory (LO, NLO, NNLO), using the 
MRST PDF sets we plot three curves, corresponding to 
\begin{itemize}
\item Common variation of the renormalization and factorization scales, 
$\mu_F = \mu_R = \mu$, but over a larger range of $\mu$, 
$M/5 < \mu < 5M$ (solid curves); 
\item variation of the factorization scale alone, setting 
$\mu_R = M_Z$ (dashed curves);
\item variation of the renormalization scale alone, setting 
$\mu_F = M_Z$ (dotted curves).
\end{itemize}
Because the LO result is independent of $\alpha_s(\mu_R)$, the
third curve is trivially constant at LO, and the former two LO curves 
lie on top of each other.   We can see from Fig.~\ref{LHCZy0mu} that 
the tiny NNLO scale variation in Fig.~\ref{LHC_Z_Mz} is not peculiar
to the range $M/2 < \mu < 2M$ used there.  Even extending the range
to $M/5 < \mu < 5M$, for a common variation the bandwidth only 
enlarges from 0.5\% to 1.2\%.   Over this same range, holding $\mu_F$
fixed and varying $\mu_R$ also produces a quite small range of
values, less than 0.5\%.  The largest variations are found by holding 
$\mu_R$ fixed and varying $\mu_F$.  These variations are still only of
order 0.7\% over the range $M/2 < \mu < 2M$, but rise to of order
5\% at the ends of the extended range $M/5 < \mu < 5M$.  The latter
are fairly extreme scale choices, however.  We believe that the range used 
in the rest of the paper, $\mu_F = \mu_R = \mu$ and $M/2 < \mu < 2M$,
provides a good guide to the perturbative uncertainty remaining from the 
terms beyond NNLO.

\noindent

\begin{figure}[htbp]
\vspace{0.0cm}
\centerline{
\psfig{figure=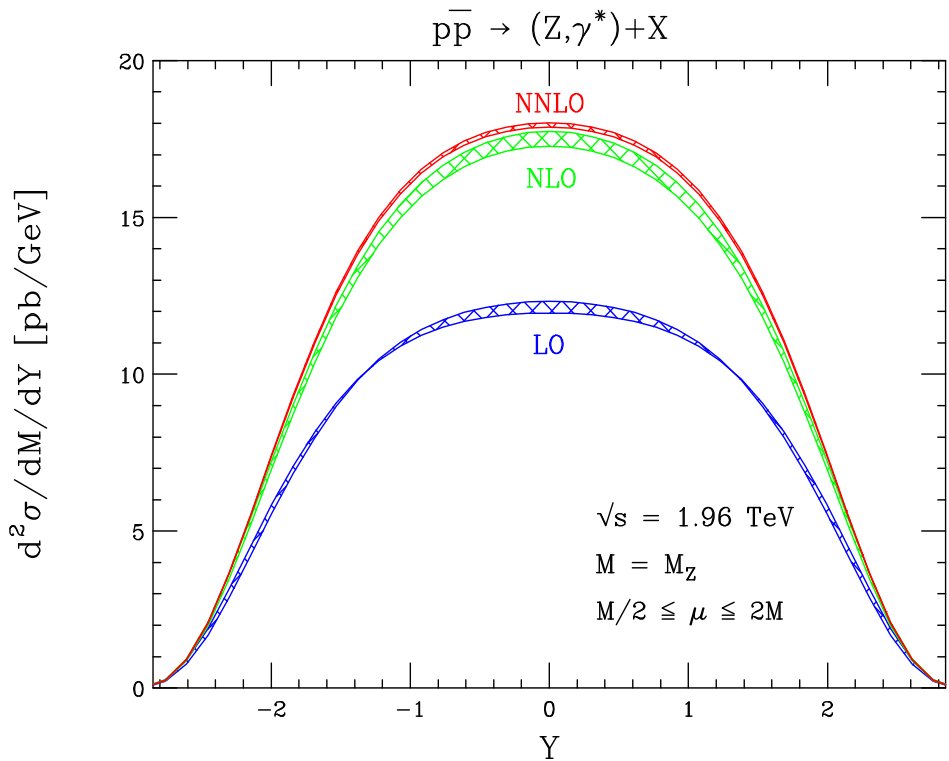,height=10.0cm,width=12.5cm,angle=0}}
\caption{The CMS rapidity distribution of an on-shell $Z$ boson at Run II of
the Tevatron.  The LO, NLO, and NNLO results have been included.  
The bands indicate the variation of the renormalization and factorization 
scales in the range $M_Z/2 \leq \mu \leq 2M_Z$.}
\label{TEVII_Z_Mz}
\end{figure}

In Fig.~\ref{TEVII_Z_Mz} we present the rapidity distribution for
on-shell $Z$ production at Run II of the Tevatron.  
The scale variation is unnaturally small at LO;
it is 3\% at central rapidities, and varies from 0.1\% to 5\% from $Y=1$
to $Y=2$.  This occurs because the direction of the scale variation
reverses within the range of $\mu$ considered, {\it i.e.}, 
$d\sigma_{LO}/d \mu = 0$ for a value of $\mu$ which satisifes 
$M_Z/2 \leq \mu \leq 2M_Z$.  This value of $\mu$ depends upon rapidity, 
leading to scale dependences which vary strongly with $Y$.  The scale variation
exhibits a more proper behavior at NLO, starting at 3\% at central
rapidities and increasing to 5--6\% at $Y=2.5$.  At NNLO the scale
dependence is drastically reduced, as at the LHC, and remains below 1\%
for all relevant rapidity values.  The magnitude of the higher-order
corrections is slightly larger at the Tevatron than at the LHC.  The NLO
prediction is higher than the LO result by nearly 45\% at central
rapidities; this shift decreases to 30\% at $Y=1.5$ and to 15\% at
$Y=2.5$.  The NNLO corrections further increase the NLO prediction by
3--5\% over the rapidity range $Y \leq 2$.

\noindent

This remarkable stability of the rapidity distribution with respect to
scale variation cannot be attributed to the smallness of the NNLO QCD
corrections to the partonic cross sections.  These corrections are the
$d\sigma^{(2)}/dY$ terms defined in Eq.~(\ref{eq:bareexpansion})
(after renormalization and mass factorization),
convoluted with the MRST PDFs and with all partonic channels included.  We
vary the scale in these terms, and normalize this variation to the NLO
cross section.  We find that the NNLO corrections contribute a scale
dependence of $\approx 5\%$ at central rapidities.  When we form the
complete NNLO cross section, which requires adding these corrections to
the convolution of the $d\sigma^{(0)}/dY$ and $d\sigma^{(1)}/dY$ terms of
Eq.~(\ref{eq:bareexpansion}) with NNLO PDFs, the width of this band is
decreased to less than 1\%.  This demonstrates a remarkable interplay
between NNLO calculations and parton distribution functions.

\begin{figure}[htbp]
\vspace{0.0cm}
\centerline{
\psfig{figure=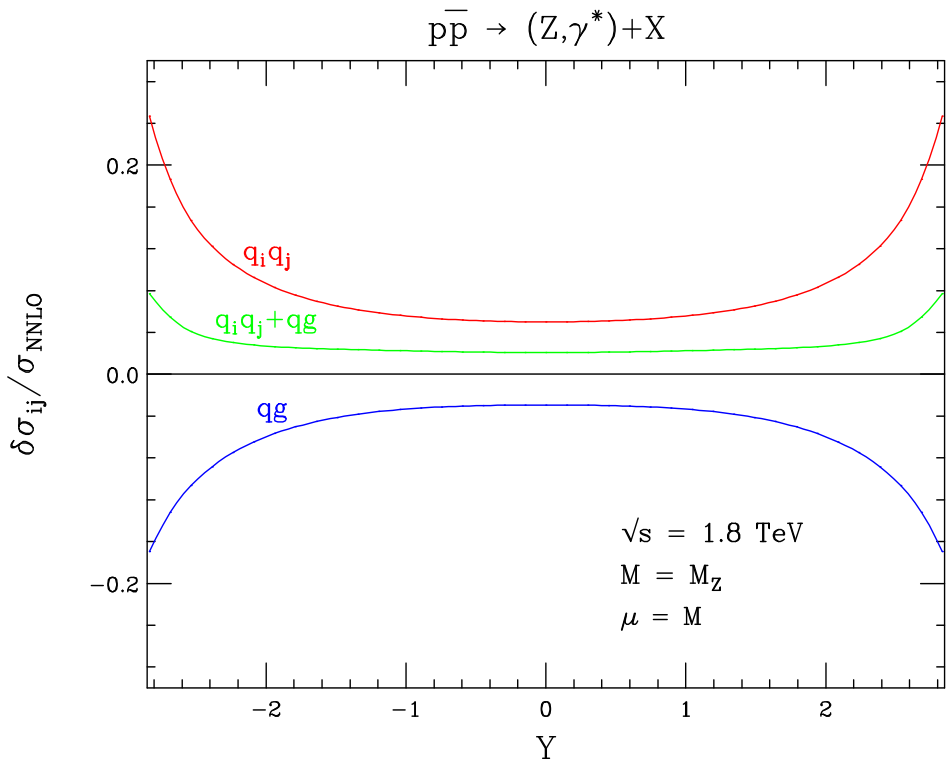,height=10.0cm,width=12.5cm,angle=0}}
\caption{The fractional contribution of the various NNLO partonic channels
to the entire NNLO cross section for $Z$ production at Run I of the Tevatron for
$\mu=M_Z$.  $q_iq_j$ denotes all quark-quark and quark-antiquark channels,
while $qg$ indicates the quark-gluon and antiquark-gluon subprocesses.
The $gg$ channel is numerically small, and would be consistent with zero
on this plot.}
\label{TEV_Z_Mz_part}
\end{figure}

The small size of the NNLO corrections is partly due to large
cancellations between the various partonic channels.  To illustrate this,
we present in Fig.~\ref{TEV_Z_Mz_part} the fractional contributions of the
various NNLO partonic corrections to the entire NNLO cross section, at Run
I of the Tevatron.  We include the $qg$ and $q_iq_j$ channels (the latter
includes $qq$ and $q\bar{q}$ inital states); the $gg$ subprocess is numerically
unimportant in this process.  The magnitude of each order $\alpha_s^2$ 
partonic correction, $\delta\sigma_{ij}$, can be 7--8\% of the complete 
NNLO cross section, $\sigma_{\rm NNLO}$, at central rapidities, 
and can reach 10\% of the entire result at larger rapidities.
They cancel significantly, however, and their sum is only $\approx 3\%$ of
the NNLO result.  This cancellation is even larger at LHC energies; in
fact, the $q_iq_j$ and $qg$ channels cancel to such an extent that the $gg$
subprocess becomes an important contribution to the NNLO corrections.
This split into partonic components is admittedly not entirely physical,
as they are linked by initial-state collinear singularities.  However,
this degree of cancellation should be rather sensitive to the PDF set
chosen.  A different choice of PDFs may lead to changes in the cross
section that are larger than that found by varying the renormalization
and factorization scales.

\noindent

\begin{figure}[htbp]
\vspace{0.0cm}
\centerline{
\psfig{figure=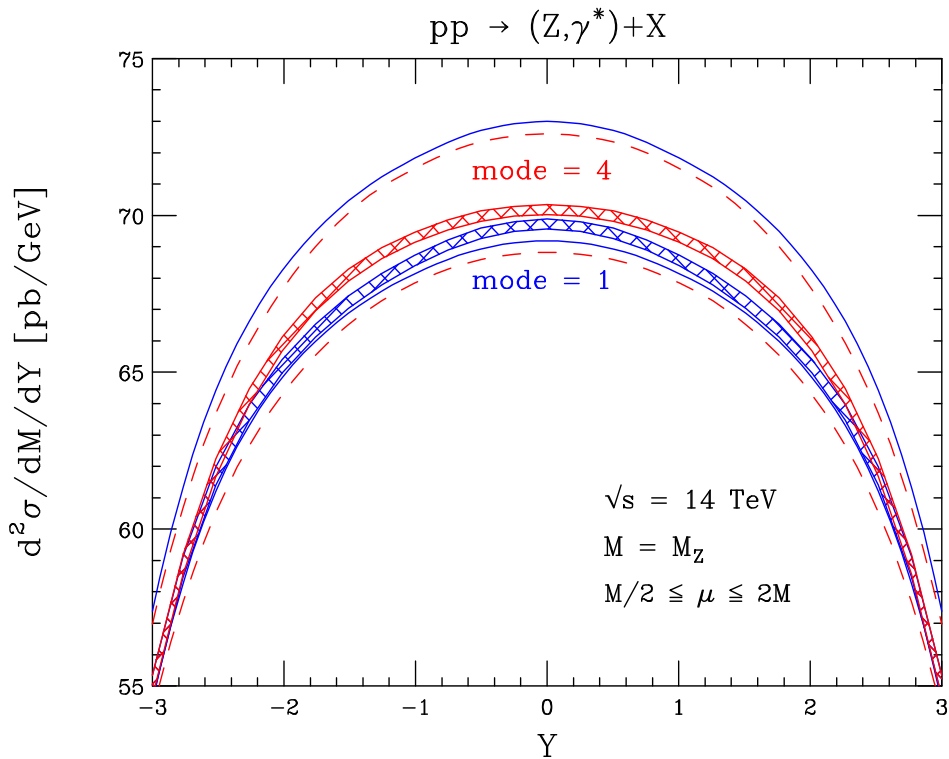,height=10.0cm,width=12.5cm,angle=0}}
\caption{The rapidity distributions for $Z$ production at the LHC for 
the MRST PDF sets {\sf mode} 1 and {\sf mode} 4.  The bands indicate 
the NNLO scale dependences, the solid lines denote the NLO {\sf mode} 1 
scale depedence, and the dashed lines indicate the NLO {\sf mode} 4 scale 
variation.  The upper lines correspond to the scale choice $\mu=2M$ 
in the NLO cross sections, while the lower lines indicate $\mu=M/2$.}
\label{LHC_Z_Mz_pdfs}
\end{figure}

To investigate how the choice of PDFs affects the NNLO cross section, we
first vary the MRST {\sf mode}.  The choices corresponding to the fast and
slow DGLAP evolutions produce negligible shifts in our result, much less
than 1\% for all rapidities studied, and smaller than the residual scale
dependence.  (Similar results have been observed at the level of
the total cross section~\cite{Martin2002}.)
However, the choice of MRST {\sf mode} 4, which provides a
better fit to the Tevatron high-$E_T$ jet data, shifts the NNLO $Z$
production cross section significantly.  We present in
Fig.~\ref{LHC_Z_Mz_pdfs} the rapidity distributions for LHC $Z$ production
using these two PDF choices.  Both the NLO and NNLO results have been
displayed; the scale variations are also included.  The two {\sf mode}
choices are indistinguishable at NLO, due to the large residual scale
dependence.  At NNLO they become quite distinct, and the $\approx 1\%$
discrepancy is potentially visible given projected LHC errors.  We note
that the difference between the two PDF sets does not just produce a shift
in the overall normalization.  The {\sf mode} 4 set slightly increases 
the number of quarks at $x \sim 0.03$, and decreases the number
of gluons more substantially in this $x$ range (to compensate for an 
even larger increase in $g(x)$ at very large $x$).  The $qg$ channel 
has a negative partonic cross section; thus, paradoxically, 
decreasing $g(x)$ increases the gluonic contribution to the cross 
section.   The quark and gluon distribution shifts, plus a
2\% increase in $\alpha_s(M_Z)$, work in concert to increase the 
{\sf mode} 4 predictions, relative to {\sf mode} 1, for $Z$ production at 
the LHC at central rapidities, and particularly in the range $1 < Y < 2$.

\noindent

\begin{figure}[htbp]
\vspace{0.0cm}
\centerline{
\psfig{figure=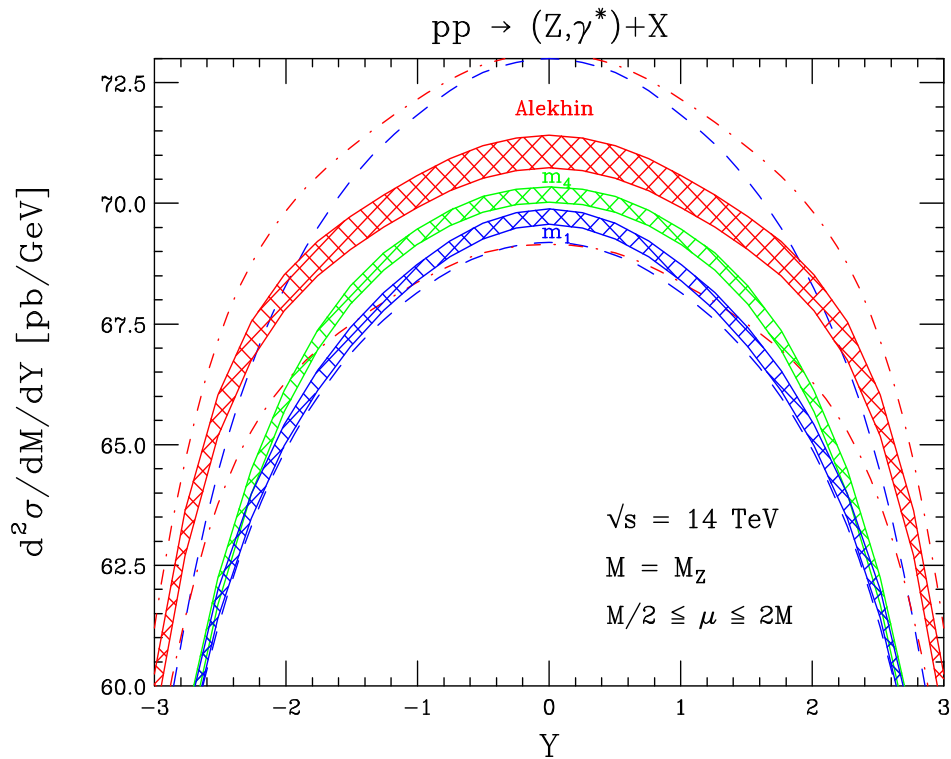,height=10.0cm,width=12.5cm,angle=0}}
\caption{The rapidity distributions for $Z$ production at the LHC for the
MRST PDF sets {\sf mode} 1 and {\sf mode} 4, and for the Alekhin PDF set.
The bands indicate the NNLO scale dependences; ${\rm m}_1$ denoted the
MRST {\sf mode} 1 set, while ${\rm m}_4$ indicates the MRST {\sf mode} 4
set.  The dashed lines denote the NLO scale dependence for the {\sf mode}
1 set, and the dot-dashed lines denote the NLO scale dependence for the
Alekhin set.  The upper lines correspond to the scale choice $\mu=2M$ in
the NLO cross sections, while the lower lines indicate $\mu=M/2$. }
\label{LHC_Z_Mz_pdfs_al}
\end{figure}

Another set of PDFs extracted with NNLO precision has been presented by
Alekhin~\cite{Alekhin2002}.  Only deep inelastic scattering data is
used in this extraction; the NNLO QCD corrections can therefore be
consistently included.  The MRST global fits utilize processes for which
these corrections are not known.  This introduces an additional source of
theoretical uncertainty into these parameterizations which is difficult to
quantify.  We present in Fig.~\ref{LHC_Z_Mz_pdfs_al} a comparison between
the MRST and Alekhin PDF sets for resonant $Z$ production at the LHC.  We
have included the NNLO scale dependences for the Alekhin set and for the
MRST {\sf mode} 1 and {\sf mode 4} sets; the NLO scale dependences for the
MRST {\sf mode} 1 and Alekhin parameterizations are also displayed.  The
large scale dependences again render all three choices indistinguishable
at NLO.  However, significant discrepancies appear at NNLO.  The
difference between the {\sf mode} 1 and Alekhin sets is 2\% at central
rapidities; this increases to 4.5\% at $Y=2$ and to 8.5\% at $Y=3$.  The
discrepancies in both normalization and shape will be clearly resolvable at
the LHC.  Although the MRST {\sf mode} 4 choice is closer to both the 
shape and normalization of the Alekhin set, the differences still
range from 1--8.5\% as the rapidity is increased; this will again be
observable at the LHC.  Electroweak gauge boson production becomes a
powerful discriminator between different PDF parameterizations when the
NNLO QCD corrections are included.

\noindent

\begin{figure}[htbp]
\vspace{0.0cm}
\centerline{
\psfig{figure=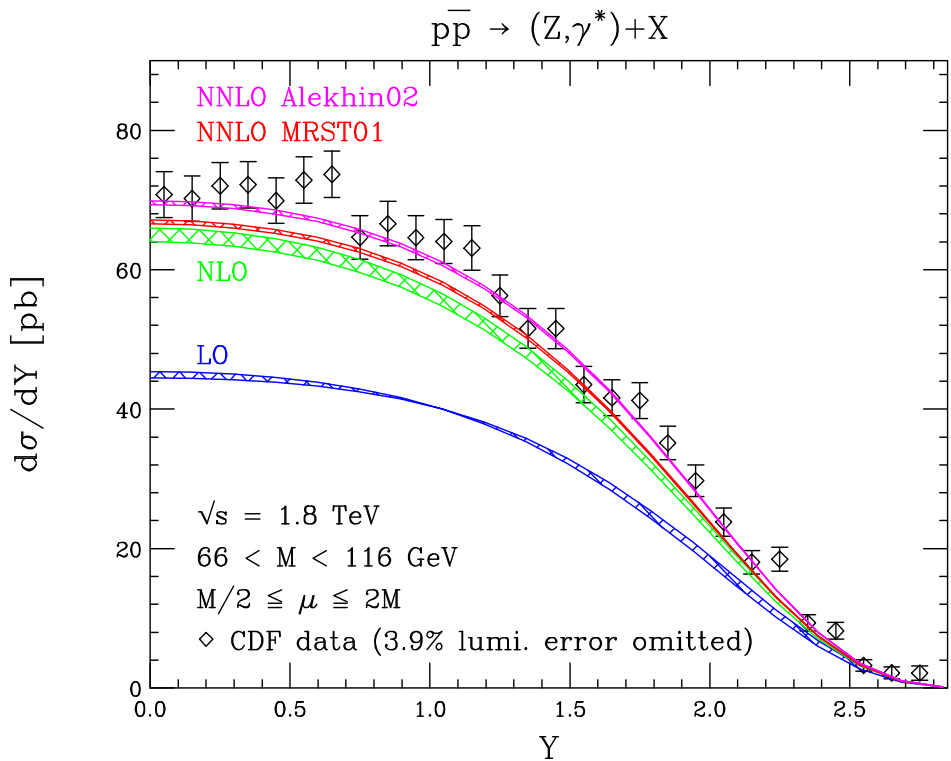,height=10.0cm,width=12.5cm,angle=0}}
\caption{The di-lepton rapidity distribution for $(Z,\gamma^*)$ production at 
Run I of the Tevatron, compared with data from CDF~\cite{ZrapCDF}.
The LO and NLO curves are for the MRST PDF set.  The thin NNLO bands
are for the MRST (lower) and Alekhin (upper) parameterizations.
The bands correspond to varying $M/2 \leq \mu \leq 2M$.}
\label{TEV_Z_Mz_DATA}
\end{figure}

The di-lepton rapidity distribution for $(Z,\gamma^*)$ production has been
measured by CDF at Run I of the Tevatron, in a mass window around $M_Z$,
$66 < M < 116$~GeV~\cite{ZrapCDF}.  To compare with these data, we
numerically integrate over $M$ as well as $z$ and $y$ in
Eq.~(\ref{zyintegral}).  The result is shown in Fig.~\ref{TEV_Z_Mz_DATA}.
(The result of doing this $M$ integral in a narrow-resonance
approximation, taking into account the finite-mass endpoints, but
neglecting photon exchange, is about 2\% lower.)  The result with the
Alekhin PDF set is about 4--5\% above the MRST result.  Naively, the
Alekhin set gives a better fit to the data.  However, most of the
Alekhin/MRST difference here is in the overall normalization, and there is
a 3.9\% overall normalization uncertainty in the data (not shown in the
error bars) due to the $p\bar{p}$ luminosity uncertainty.  Also,
electroweak corrections have not yet been included.  Hence the two PDF
sets probably cannot be distinguished by this Run I data.  Instead, it is
clear from the figure that, for a given PDF set, the di-lepton rapidity
distribution around the $Z$ mass may be used to `monitor' the luminosity
at Run II, for which the statistical errors will be significantly smaller
than those shown.

\noindent

\begin{figure}[htbp]
\vspace{0.0cm}
\centerline{
\psfig{figure=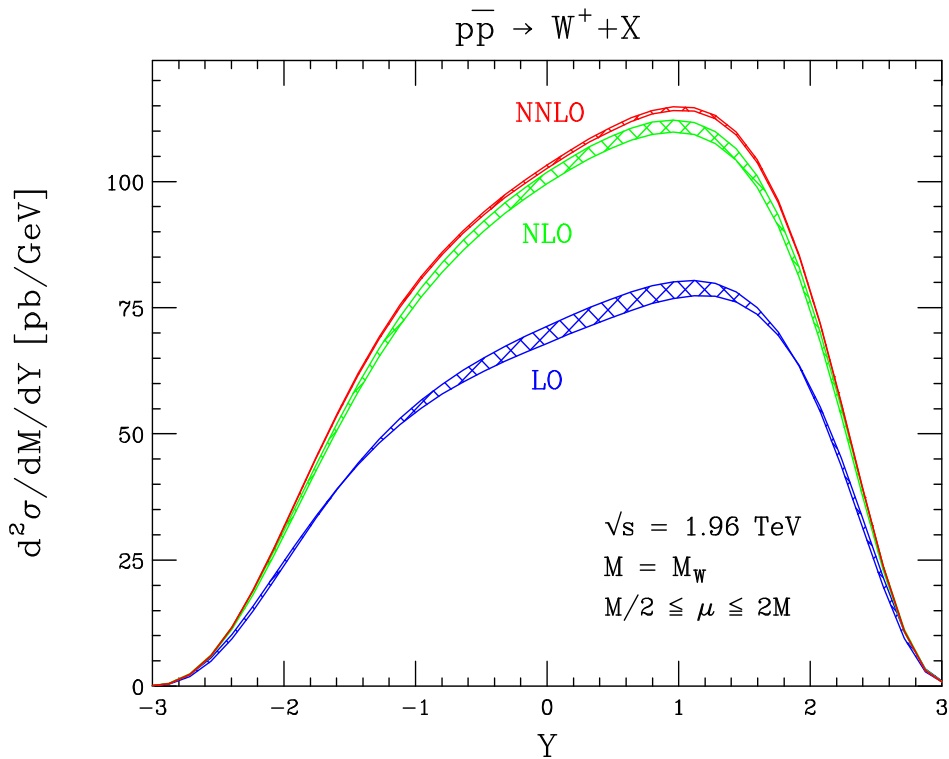,height=10.0cm,width=12.5cm,angle=0}}
\caption{The CMS rapidity distribution of an on-shell $W^+$ boson at Run
II of the Tevatron.  Shown are the LO, NLO, and NNLO results for the MRST
PDF sets.  The bands indicate the variation of the renormalization and 
factorization scales in the range $M_W/2 \leq \mu \leq 2M_W$.}
\label{TEVII_Wp_Mw}
\end{figure}

We now examine the resonant production of $W$ bosons at Run II of the
Tevatron.  We present in Fig.~\ref{TEVII_Wp_Mw} the rapidity distribution
for $W^+$ production; the distribution for the $W^-$ can be obtained by
substituting $Y \rightarrow -Y$.  Both the scale variations and the
magnitudes of the higher corrections are similar to those found previously
for $Z$ production at the Tevatron.  The scale dependence at LO is again
unnaturally small, ranging from 3--5\%, because $d \sigma_{LO}/d \mu = 0$
for values of $\mu$ within the parameter space studied.  At NLO the scale
variations are between 2\% and 3.5\%; they decrease to $\approx 0.3$--0.7\% at
NNLO, depending upon the rapidity chosen.  The magnitude of the NLO
corrections is large, varying from 45\% at central rapidities to 
$\approx 25\%$ at larger rapidities.  The NNLO corrections are also 
appreciable; they range from 2.5\% at $Y=0$ to 4\% at $|Y| \approx 2$.

\noindent

\begin{figure}[htbp]
\vspace{0.0cm}
\centerline{
\psfig{figure=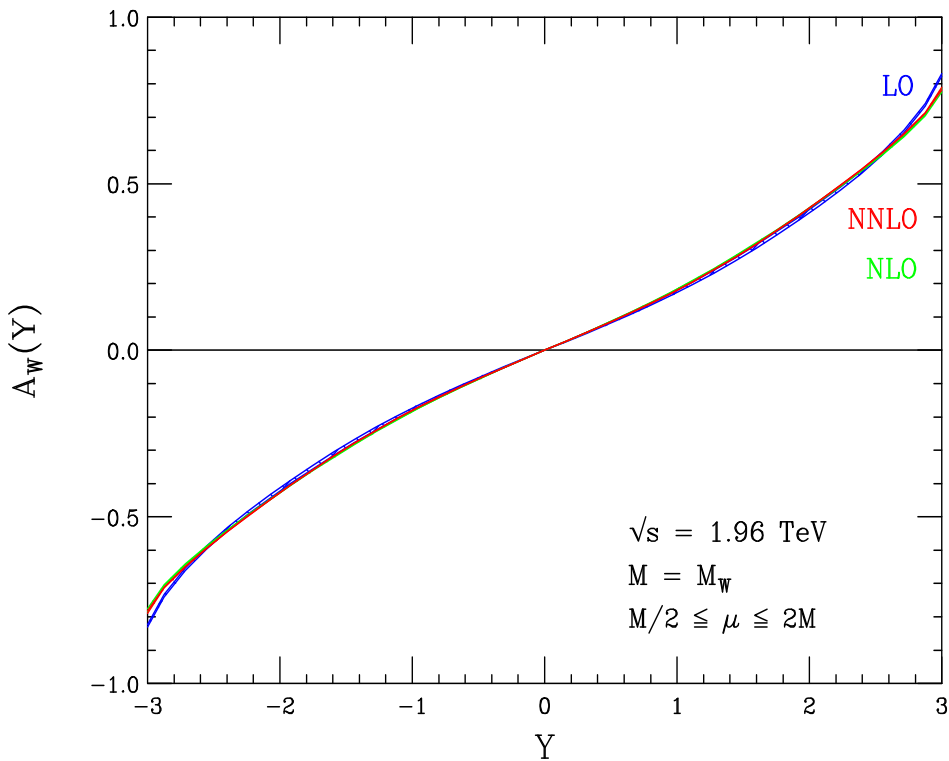,height=10.0cm,width=12.5cm,angle=0}}
\caption{The $W$ charge asymmetry at Run II of the Tevatron.  Included 
are the LO, NLO, and NNLO results.  The bands indicate the variation of the
renormalization and factorization scales in the range $M_W/2 \leq \mu \leq
2M_W$.  As the charge asymmetry is rather insensitive to QCD corrections,
the three bands are almost completely degenerate.}
\label{TEVII_ca_Mw}
\end{figure}

Another observable frequently studied at hadron colliders is the 
$W$ charge asymmetry, defined as 
\begin{equation}
A_W(Y) = 
\frac{d\sigma(W^+)/dY - d\sigma(W^-)/dY}
{d\sigma(W^+)/dY + d\sigma(W^-)/dY}.
\label{eq:W_chargeasym}
\end{equation}
A simple calculation in the LO approximation reveals that this quantity is
sensitive to the $x$ dependence of $u(x)/d(x)$, the ratio of up and down
quark distributions in the proton.  Although in a realistic experiment
only the pseudorapidity of the charged lepton coming from the $W$ decay
can be measured, much of the sensitivity to the PDFs remains.  Since $A_W$
is a ratio of cross sections, it might be expected that it is rather
insensitive to QCD corrections.  This is indeed the case.  
At the Tevatron, a $p\bar{p}$ collider, with the assumption of CP 
invariance, the charge asymmetry is an odd function of $Y$, since it 
may be written as
\begin{equation}
A_W(Y) = 
\frac{d\sigma(W^+)/dY - d\sigma(W^+)/dY |_{Y\to-Y}}
{d\sigma(W^+)/dY + d\sigma(W^+)/dY |_{Y\to-Y}}
= - A_W(-Y).
\label{eq:W_chargeasym_ppbar}
\end{equation}
The asymmetry is positive for positive $Y$, corresponding to the 
$W^+$ boson moving in the same direction as the incident proton,
because $u(x)$ is larger than $d(x)$ at large $x$.
In Fig.~\ref{TEVII_ca_Mw}, we present the LO, NLO, and NNLO predictions
for the charge asymmetry at Run II of the Tevatron, together with their
scale dependences.  The NLO corrections increase the Born level result by
2--4\%.  The NNLO corrections to the NLO result range from $-2$\% at
central $W$ rapidities to $+1$\% at large rapidities.

\begin{figure}[htbp]
\vspace{0.0cm}
\centerline{
\psfig{figure=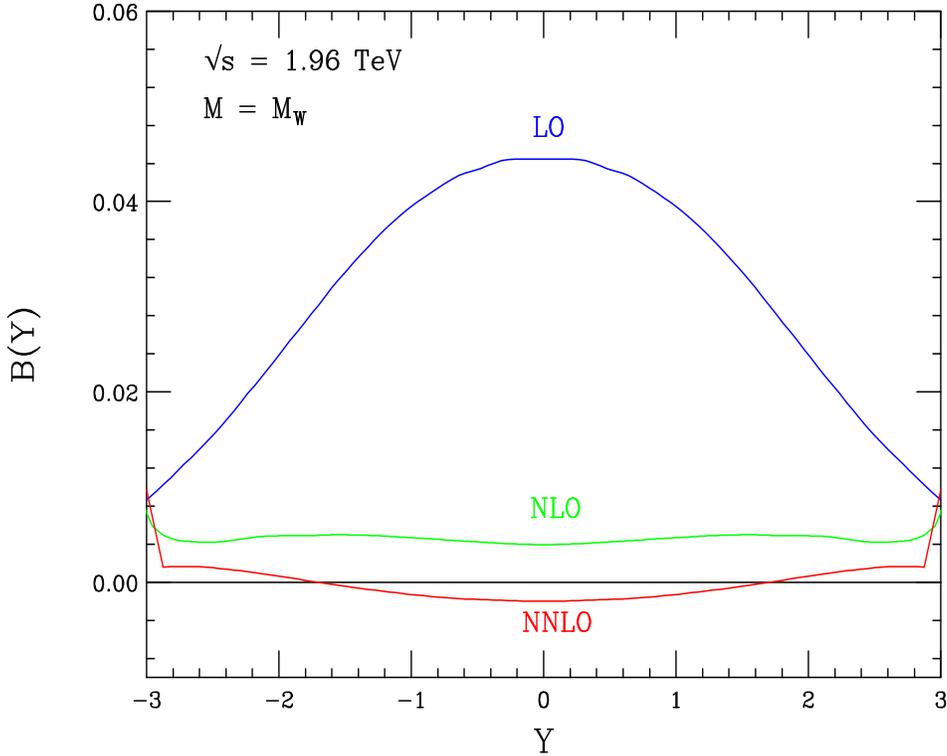,height=10.0cm,width=12.5cm,angle=0}}
\caption{The scale-dependence bandwidths for the $W$ charge asymmetry at
the Tevatron.  Included are the LO, NLO, and NNLO results.}
\label{TEVII_cabw_Mw}
\end{figure}

The scale variations of $A_W(Y)$ are small; to study them, we
present in Fig.~\ref{TEVII_cabw_Mw} the scale-dependence bandwidths,
defined as
\begin{equation}
B(Y) = \frac{A_W(Y,\mu=2M_W)-A_W(Y,\mu=M_W/2)}{A_W(Y,\mu=M_W)}.
\end{equation}
The scale variation is already below 5\% for all rapidities at LO, and is
below 1\% at NLO.  The NNLO prediction is absolutely stable against scale
variation, indicating that this observable is potentially a very strong
constraint on quark distribution functions.  We note that the scale choice
$\mu=2M_Z$ in the LO asymmetry yields an approximation to the NNLO result
which is accurate to 1--2\% for essentially all rapidities.

\noindent

\noindent

\begin{figure}[htbp]
\vspace{0.0cm}
\centerline{
\psfig{figure=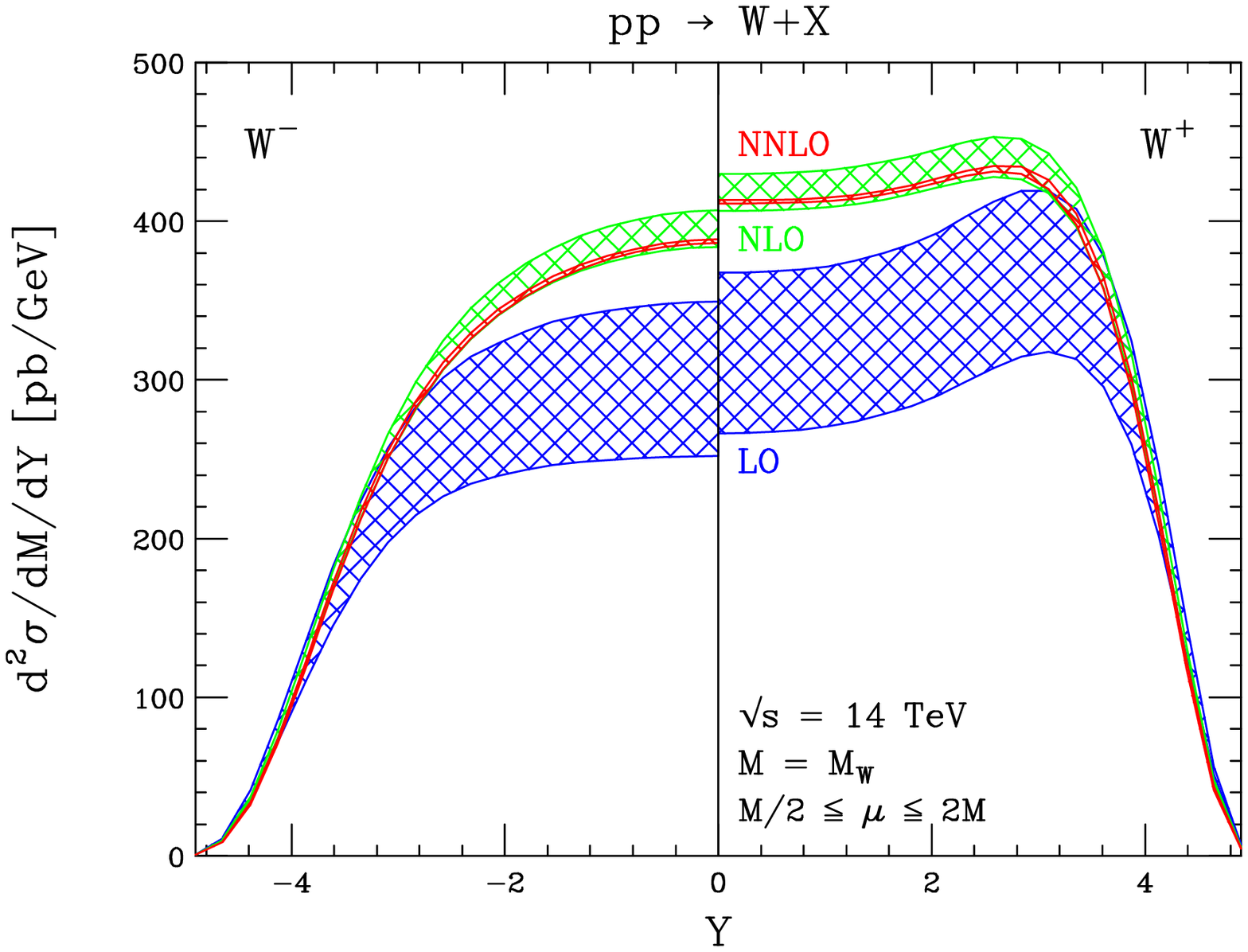,height=10.0cm,width=12.5cm,angle=0}}
\caption{The CMS rapidity distributions for production of an on-shell
$W^-$ boson (left) and on-shell $W^+$ boson (right) at the LHC,
at LO, NLO, and NNLO, for the MRST PDF sets.  Each distribution
is symmetric in $Y$; we only show half the rapidity 
range in each case.  The bands indicate the common variation of the 
renormalization and factorization scales in the range 
$M_W/2 \leq \mu \leq 2M_W$.}
\label{LHC_W_Mw}
\end{figure}

The NNLO predictions for the rapidity distributions for on-shell
$W^\pm$ boson production at the LHC are shown in Fig.~\ref{LHC_W_Mw}.
The distributions are symmetric in $Y$; only the positive half
of the rapidity range is shown for $W^+$, and the negative half
for $W^-$.  The charge asymmetry is positive
for all rapidities, but is particularly striking around $Y = 3$.
The behavior of the perturbation series is very
similar to that discussed previously for $Z$ production at the LHC.
Again the NNLO scale-variation bandwidths are extremely narrow
for central rapidities, ranging from $\approx0.6$\% for $Y < 2$,
to 1.5\% at $Y=3$, to 3\% at $Y=4$.

\noindent

\begin{figure}[htbp]
\vspace{0.0cm}
\centerline{
\psfig{figure=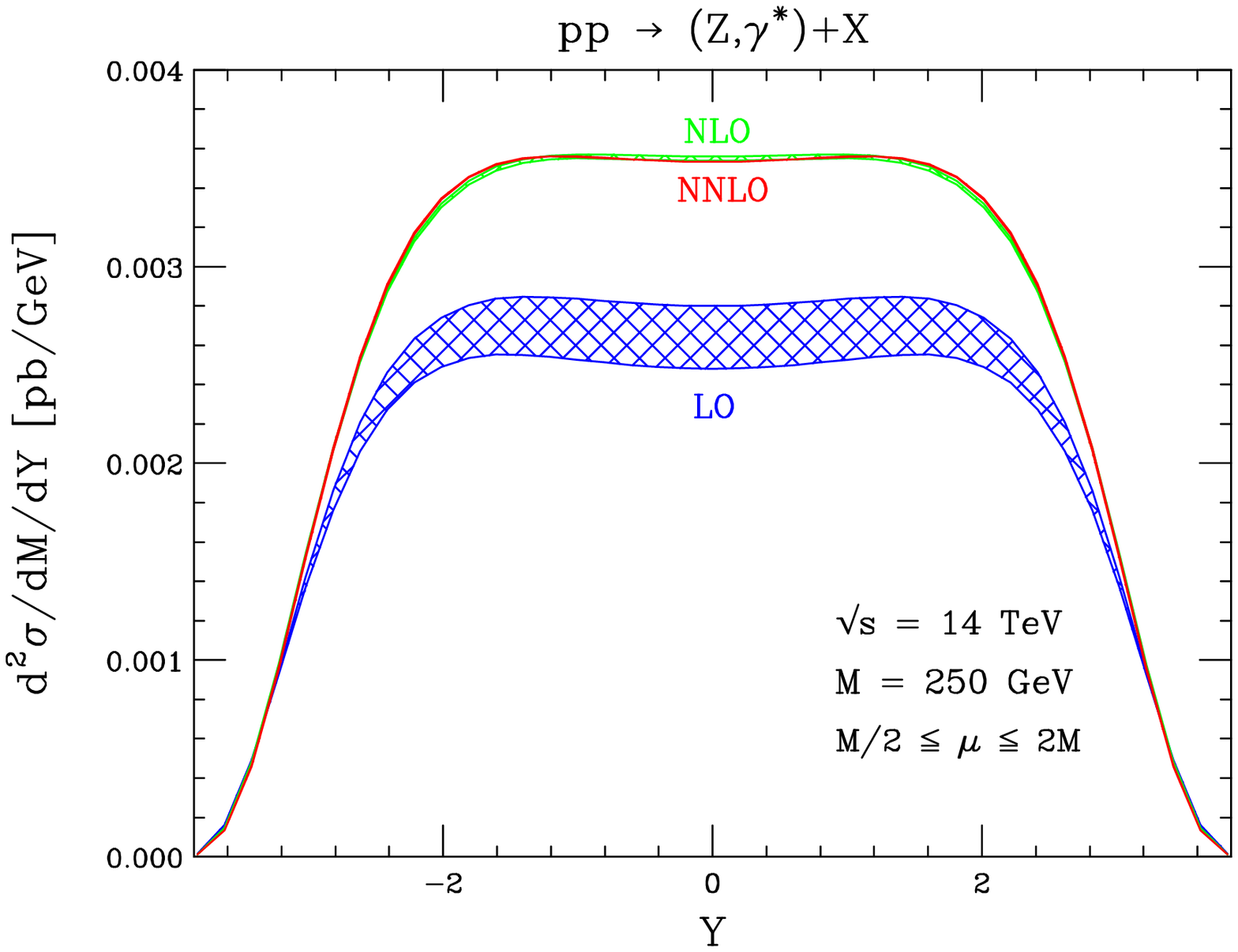,height=10.0cm,width=12.5cm,angle=0}}
\caption{The rapidity distribution for $(Z,\gamma^*)$ production at the
LHC for an invariant mass $M = 250$ GeV.  The LO, NLO, and NNLO results
have been included.  The bands indicate the residual scale dependences.}
\label{LHC_Z_250}
\end{figure}

\begin{figure}[htbp]
\vspace{0.0cm}
\centerline{
\psfig{figure=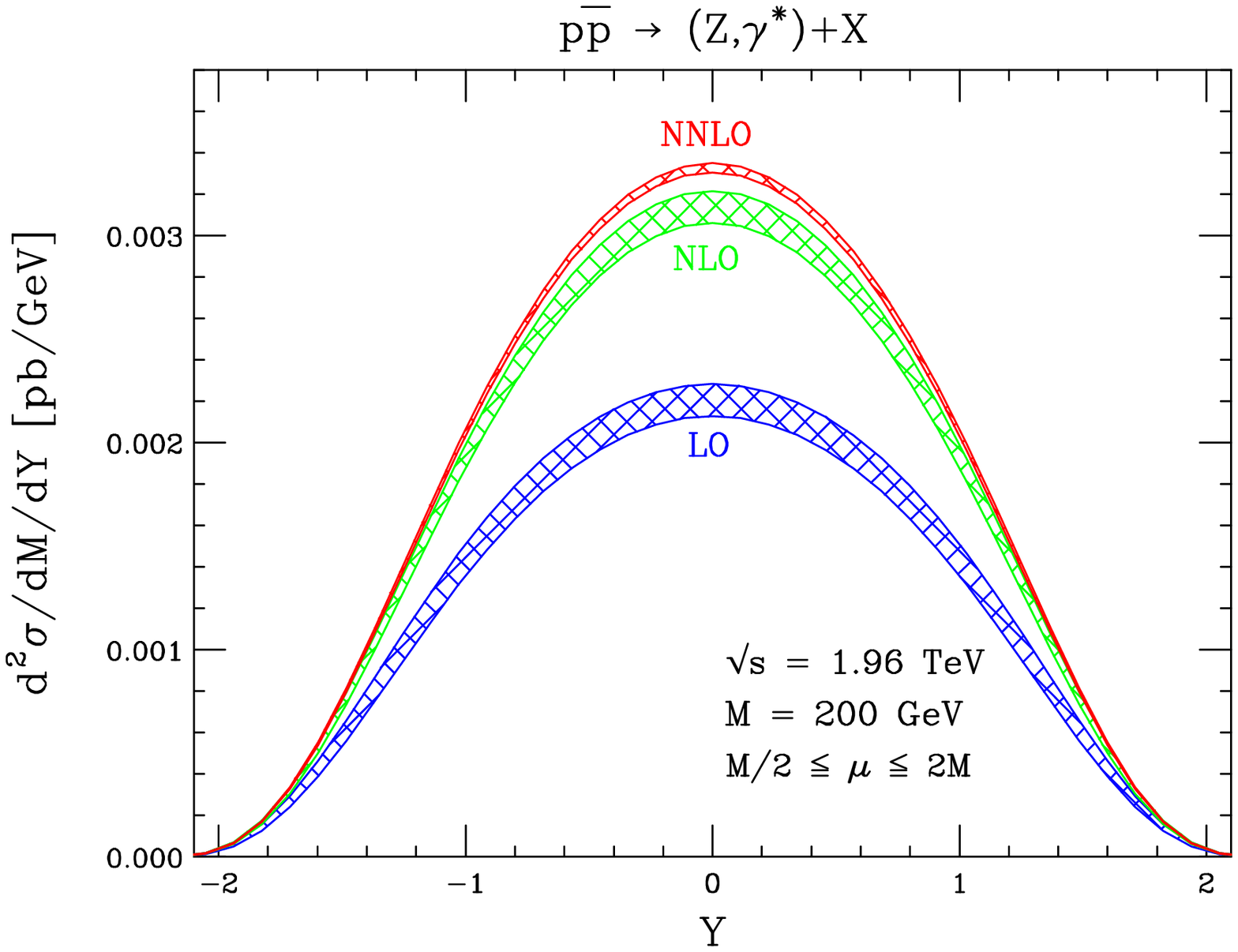,height=10.0cm,width=12.5cm,angle=0}}
\caption{The rapidity distribution for $(Z,\gamma^*)$ production at Run II
of the Tevatron for an invariant mass $M = 200$ GeV.  The LO, NLO, and
NNLO results have been included.  The bands indicate the residual scale
dependences.}
\label{TEVII_Z_200}
\end{figure}

In addition to the study of resonant production of electroweak gauge
bosons, both the Tevatron and the LHC use high-invariant-mass Drell-Yan
production of lepton pairs to search for new gauge bosons and lepton-quark
contact interactions.  Although these are primarily inclusive searches,
rapidity cuts are required because of experimental constraints.  We
therefore examine the NNLO QCD corrections to off-shell $(Z,\gamma^*)$
production at large invariant masses.  We present below the rapidity
distribution for $M=250$ GeV $(Z,\gamma^*)$ production at the LHC in
Fig.~\ref{LHC_Z_250}, and for $M=200$ GeV at Run II of the Tevatron in
Fig.~\ref{TEVII_Z_200}.  The scale dependences are significantly smaller for
$M=250$ GeV than for resonant $Z$ production at the LHC.  The LO scale
variation is 12\% at central rapidities and 4\% at $Y=3$.  Both the NLO
and NNLO scale variations are much less than 1\% for all values of
rapidity.  The magnitude of the higher-order corrections is much larger,
however.  The NLO result increases the LO prediction by nearly 35\% at
central rapidities; this correction decreases to 10\% at larger $Y$
values.  This discrepancy between the sizes of the scale variations and of
the NLO shifts sends a somewhat mixed message regarding the importance of
the NNLO corrections.  We find that they are small, decreasing the NLO
result by less than 0.5\% for $Y < 1.5$, and increasing it by less than 
1\% for $1.5 < Y < 2.8$.  The small scale dependence of the NNLO cross 
section and the stability of the NLO prediction indicate a complete 
stabilization of the perturbative result for $M=250$ GeV at the LHC.

The results for $M=200$ GeV $(Z,\gamma^*)$ production at Run II of the
Tevatron exhibit both larger scale dependences and more important 
higher-order corrections.  The LO scale variations are similar to those 
found at the LHC, ranging from 7\% at $Y=0$ to $\approx 15\%$ at larger
rapidity values.  In contrast to the LHC case, the NLO scale dependences 
remain fairly large, varying from 5\% at central rapidities to 14\% 
at $Y=2$.  At NNLO, the scale variations are between 1.5\% and 4\%, 
again increasing for larger rapidities.  The magnitude of the NLO 
corrections is over 40\% at central rapidities, and $\approx 30\%$ 
at larger $Y$ values.  The NNLO corrections further increase the 
NNLO result by 5--6\% throughout the entire rapidity range.

\noindent

\noindent

Finally, we study the accuracy of various approximations to the complete
NNLO correction to the rapidity distribution.  There are three distinct
types of terms which appear in the result: 
\begin{itemize}
\item {\it soft} ($s_z$): terms which contain either
a delta function or a plus distribution in $1-z$.
These terms arise from production of the vector boson $V$ close to the 
partonic threshold, and can be obtained by considering only soft partonic 
emissions from the $q\bar{q} \rightarrow V$ subprocess.
\item {\it collinear} ($c_y$):  terms containing delta
functions or plus distributions in either $y$ or $1-y$, but not in $1-z$.
These terms result from the emission of radiation collinear to one of 
the initial partons.
\item {\it hard} ($h$): terms which have no delta functions or plus 
distributions.  These terms arise from generic scattering events with 
the emission of hard additional partons in the final state. 
\end{itemize}
There is some potential ambiguity in this separation, due to the presence
of Jacobian factors in the integration.  We perform the separation
in terms of the functions $F_{ij}(z,y)$ appearing in Eq.~(\ref{zyintegral});
{\it i.e.}, including all Jacobian factors resulting from the
transformation the variables $(z,y)$.
The $s_z$ terms can be obtained by using the soft gluon
approximation, and it is possible to imagine obtaining the $c_y$
contributions from a simplified calculation in which the collinear
emission of $V$ is factorized from a hard scattering piece.  The hard
emissions, however, require a full NNLO computation.  Intuitively, we
expect the $s_z$ terms, which are the simplest to obtain, to dominate for
large invariant masses, {\it i.e.}, as the $z \rightarrow 1$ threshold is
approached.  We wish to examine whether this contribution, or perhaps the
$s_z$ and $c_y$ terms together, can furnish a reasonable approximation in
phenomenologically interesting regions of parameter space.

\begin{figure}[htbp]
\vspace{0.0cm}
\centerline{
\psfig{figure=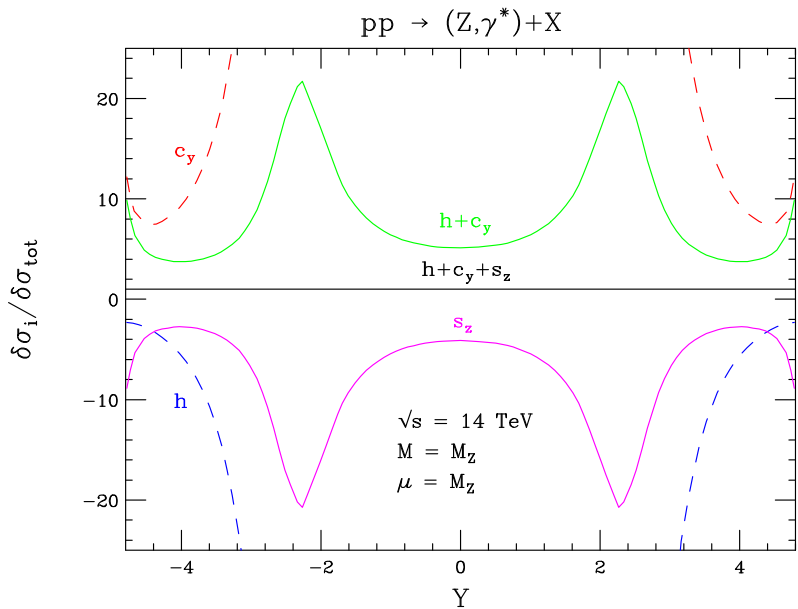,height=8.0cm,width=10.5cm,angle=0}}
\caption{The components of the NNLO corrections to the rapidity distribution
for $(Z,\gamma^*)$ production at the LHC for $M=M_Z$.  The pieces included
are the hard part $h$, $s_z$, $c_y$, and the sum of the $h$ and $c_y$ pieces.
The complete NNLO correction $h+c_y+s_z$ is normalized to unity.
We have set $\mu=M$.}
\label{LHC_Z_sp2_Mz}
\end{figure}

\begin{figure}[htbp]
\vspace{0.0cm}
\centerline{
\psfig{figure=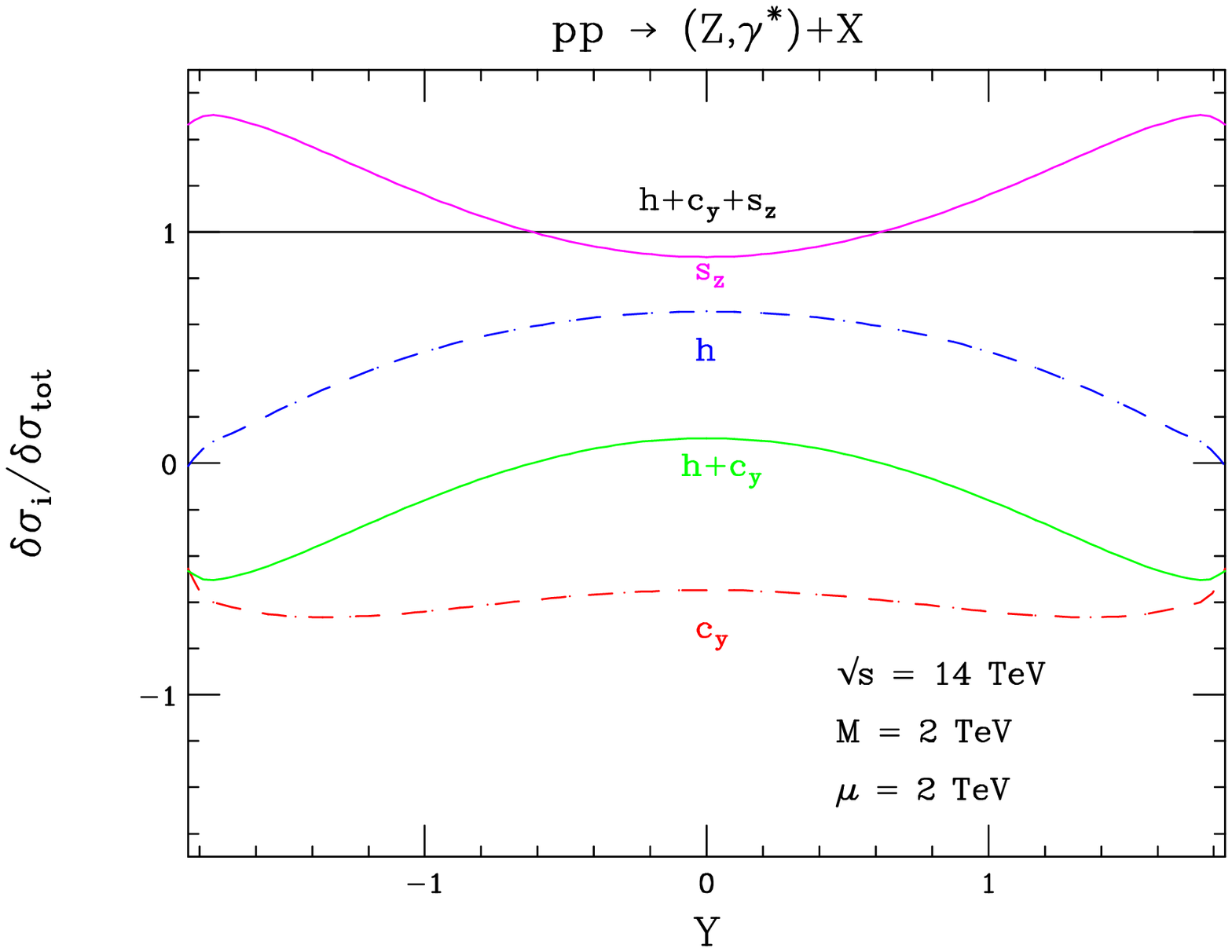,height=8.0cm,width=10.5cm,angle=0}}
\caption{The components of the NNLO corrections to the rapidity distribution
for $(Z,\gamma^*)$ production at the LHC for $M=2$ TeV.  The pieces
included are the hard part $h$, $s_z$, $c_y$, and the sum of the $h$ and
$c_y$ pieces.  The complete NNLO correction $h+c_y+s_z$ is normalized to unity.
We have set $\mu=M$.}
\label{LHC_Z_sp2_2000}
\end{figure}

We present in Figs.~\ref{LHC_Z_sp2_Mz} and~\ref{LHC_Z_sp2_2000} the NNLO
corrections to the rapidity distributions for $(Z,\gamma^*)$ production at
the LHC, split into its soft, collinear and hard components, for the 
invariant masses $M=M_Z$ and $M=2$ TeV.  The NNLO corrections are the 
$d\sigma^{(2)}/dY$ terms defined in Eq.~(\ref{eq:bareexpansion}), 
convoluted with the MRST PDFs and with
all partonic channels included.  We present separately the following
pieces: the $s_z$ term, the $c_y$ term, the $h$ term, and the sum of
the $h$ and $c_y$ pieces, which would integrate to the ``hard'' (non-soft)
part of the total cross section.  These terms are normalized to the complete NNLO
correction.  At $M=M_Z$, all components are important.  We note that there
are large cancellations between the $s_z$ term and the remaining pieces.
Neither the $s_z$ piece nor the sum of the $s_z$ and $c_y$ terms furnishes
a good approximation to the complete result.  Generic hard emissions are
important; this result is expected, since there is a large amount of 
phase space
available.  At $M=2$ TeV, the magnitude of the $s_z$ term becomes larger
compared to the hard and $c_y$ terms, as expected.  However, it still does
not furnish a good approximation to the entire result for all rapidities;
the fact that it does so for central rapidities arises from an accidental
cancellation between the hard and $c_y$ pieces.  We observe similar
behavior for Tevatron kinematics.  We note that at higher invariant
masses, the magnitude of the hard term decreases quickly.  The $c_y$ term
also decreases, but less rapidly.  The $s_z$ term does not dominate until
very large invariant masses are reached.

\noindent

\noindent

\section{Conclusions}

\label{sec:conclusions}

We have presented a calculation of the rapidity distributions for
electroweak gauge boson production at hadron colliders through NNLO in
QCD.  This is the first complete NNLO computation of a differential
quantity needed for high-energy hadron collider physics.  We have
discussed in detail a powerful new technique for calculating differential
distributions.  This method is completely automated, produces fully
analytic results, and treats the various components of a NNLO calculation
in a unified manner.  Our results will assist in the extraction of parton
distribution functions, parton-parton luminosities, electroweak gauge
boson information, and other quantities of interest with the accuracy 
needed for Tevatron and LHC physics.

We have found that the residual scale dependences for resonant $W$ and $Z$
production at both the Tevatron and the LHC are below 1\% when the NNLO
corrections are included; the rapidity distributions are completely stable
against higher-order QCD corrections.  Only higher-order electroweak
corrections and mixed QCD-electroweak effects remain to be
included~\cite{WZewkcorrs}.  These distributions are therefore ideal
observables to use to discriminate between different parton distribution
function parameterizations.  We have studied several different NNLO
extractions of parton distribution functions obtained by the MRST group,
as well as an NNLO extraction provided by Alekhin.  Varying the evolution
rate of the approximate NNLO DGLAP kernels in the MRST parameterization
yields negligible shifts in our results.  However, an MRST PDF set
designed to provide a better fit to the Tevatron high-$E_T$ jet cross
section produces a difference of about $1\%$ in rapidity distributions at
the LHC.  This difference may be observable, given expected
experimental errors.

The deviations induced by instead using Alekhin's PDF extraction are more
striking.  Both the normalization and shape of the rapidity distributions
obtained with Alekhin's parameterization differ from those found with the
MRST sets; the differences range from 2--8.5\% as the rapidity is varied.
These differences should be easily resolvable at the LHC, given the
expected errors.  The MRST parameterizations are derived from global fits
to a variety of data, including data from processes for which the NNLO QCD
corrections are unknown.  We note that the magnitude of the discrepancies
between the Alekhin and MRST PDF sets is consistent with the typical size
of NNLO QCD corrections.  It is conceivable that the inclusion of these
corrections into the MRST fit might lessen the observed differences.
In fact, the NNLO Alekhin PDF set includes a full error matrix.  (Similar
uncertainty estimates are available for the MRST set at NLO.)
This matrix permits the construction of PDF uncertainty bands for the
vector boson rapidity distributions, whereas here we just employed the
central PDF values.  We defer such a study to future work.

The magnitude of the NNLO corrections to resonant gauge boson production
ranges from 1--2\% at the LHC to 3--4\% at the Tevatron; the corrections for
higher invariant mass gauge bosons can reach 5--6\% at the Tevatron.  These
contributions must be included to yield a theoretical calculation accurate
to $\approx 1\%$, the projected experimental precision at the LHC.
However, the NNLO corrections do not vary strongly with rapidity.  The NLO
rapidity distribution appears to describe the kinematics quite well.
Reweighting the NLO distributions by the inclusive $K$-factor
$K^{(2)}=\sigma_{\rm NNLO}/\sigma_{\rm NLO}$ yields an approximation
accurate to $\leq 1\%$ for all relevant rapidities.  The analogous reweighting 
of the LO results, by $K^{\rm NNLO}=\sigma_{\rm NNLO}/\sigma_{\rm LO}$,
does {\it not} furnish a good approximation to the complete result.
The excellent accuracy of the NLO reweighting technique for the rapidity 
distribution suggests that one apply the factor $K^{(2)}$ to output 
from a hadron-level Monte Carlo program which incorporates the NLO 
vector-boson production matrix elements, such as MC@NLO 2.2~\cite{MCNLO22}.
This simple procedure should give a good picture of the structure of the 
hadronic events accompanying the vector bosons, and is likely to 
approach NNLO precision for sufficiently inclusive observables.

We have also studied the accuracy of approximating the NNLO corrections by
partial results.  We have found that including only virtual and soft gluon
corrections, labeled as $s_z$ in the text, does not yield a good
approximation for resonant gauge boson production.  Only at very large
invariant masses do these terms dominate.  We estimate that average values
of Bjorken $x \geq 0.3-0.4$ must be reached before the $s_z$ component
accounts for $\approx 80\%$ of the complete NNLO correction for all
relevant rapidities.  We also note that the $s_z$ terms do not accurately
predict the shape of the NNLO correction, as is apparent from
Figs.~\ref{LHC_Z_sp2_Mz} and~\ref{LHC_Z_sp2_2000}.

Finally, we note that with our result for the rapidity distribution, it is
possible to obtain almost full control over the kinematics of the
electroweak boson, as produced in fixed-order perturbation theory. 
This is because the NLO QCD corrections to the double
differential distribution $d^2\sigma/(dY dp_\perp)$ for electroweak boson
production are known~\cite{dpt}. It was assumed in Ref.~\cite{dpt} that
$p_\perp \ne 0$.  It is therefore not possible to perform the integration
over $p_\perp$ to get $d\sigma/dY$ using the results of Ref.~\cite{dpt}
alone. However, the NNLO calculation of the rapidity distribution
presented here gives an unambiguous answer for the integral over $p_\perp$
at fixed values of rapidity, and can therefore be used as a normalization
condition.  We write
\begin{equation}
\frac{d^2\sigma_{\rm mod}}{dY dp_\perp} = 
\theta(p_\perp - p_\perp^{\rm cut}) 
\frac{d^2\sigma}{dY dp_\perp}
+ \left [ \frac{d \sigma}{d Y} 
- \int \limits_{p_\perp^{\rm cut}}^{p_\perp^{\rm max}}
 dp_\perp
\frac{d^2\sigma}{dY dp_\perp} \right ] \theta(p_\perp^{\rm cut} 
- p_\perp  ),
\label{pt}
\end{equation}
where $d^2\sigma/(dY dp_\perp)$ is the distribution computed in 
Ref.~\cite{dpt}.  
Integrating $d^2\sigma_{\rm mod}/(dY dp_\perp)$ over $p_\perp$ 
gives the correct result for the rapidity distribution; however, 
the ``zero $p_\perp$'' bin extends from $p_\perp=0$ 
to $p_\perp = p_\perp^{\rm cut}$.  Apart from this drawback, 
Eq.~(\ref{pt}) provides a simple way to describe the electroweak boson 
kinematics at NNLO in QCD.

Our results are an important theoretical input for physics at both the 
Tevatron and the LHC.  We believe the method we have introduced to 
obtain these results can be used to calculate other phenomenologically 
interesting observables.  We anticipate its application in many other 
areas of collider physics. 

\vskip .5 cm 
\noindent
{\bf Acknowledgments}
\vskip .2 cm 
\par\noindent
We thank S. Alekhin, J. Andersen, F. Gianotti, A. Kotwal, 
W. Langeveld, M. Peskin, and W.J. Stirling for useful discussions 
and communications.  The work of K.M. is partially supported by the
DOE under grant number DE-FG03-94ER-40833 and by the Outstanding Junior
Investigator Award DE-FG03-94ER-40833.  The work of F.P. is partially 
supported by the NSF grants P420D3620414350 and P420D3620434350.


\end{document}